\documentclass[aps,prd,onecolumn,floatfix,nofootinbib,showpacs]{revtex4-1}

\pdfoutput=1
\usepackage{graphicx,color,float}
\usepackage{hyperref}
\usepackage{multirow}
\usepackage{verbatim}
\usepackage{longtable}
 \usepackage{amsmath}
\usepackage{amsfonts}
\usepackage{amssymb}
\usepackage{rotating}
\usepackage[dvipsnames]{xcolor}

\usepackage{setspace}

\usepackage{ulem}

\newcommand{\imag}{\Im {\rm m}}
\newcommand{\real}{\Re {\rm e}}
\newcommand{\lsim}{\raisebox{-0.13cm}{~\shortstack{$<$ \\[-0.07cm] $\sim$}}~}
\newcommand{\gsim}{\raisebox{-0.13cm}{~\shortstack{$>$ \\[-0.07cm] $\sim$}}~}
\def\slash#1{#1\!\!\!/}

\begin{document}

{\small
\begin{flushright}
IUEP-HEP-21-02
\end{flushright} }

\title{
Yukawa Alignment Revisited in the Higgs Basis
}

\def\slash#1{#1\!\!/}

\renewcommand{\thefootnote}{\arabic{footnote}}

\author{
Jae Sik Lee,$^{1,2}$\footnote{jslee@jnu.ac.kr}~
and
Jubin Park$^{2}$\footnote{honolov77@gmail.com}}

\affiliation{
$^1$ Department of Physics, Chonnam National University,
Gwangju 61186, Korea\\
$^2$ IUEP, Chonnam National University, Gwangju 61186, Korea 
}
\date{June 28, 2022}

\begin{abstract}
\begin{spacing}{1.30}
We implement a comprehensive and detailed study of
the alignment of Yukawa couplings in the so-called Higgs basis
taking the framework of general two Higgs doublet models (2HDMs).
We clarify the model input parameters and derive the Yukawa couplings
considering the two types of CP-violating sources: 
one from the Higgs potential and
the other from the three complex alignment parameters $\zeta_{f=u,d,e}$.
We consider the theoretical constraints from the perturbative unitarity and
for the Higgs potential to be bounded from below as well as 
the experimental ones from electroweak precision observables.
Also considered are the constraints on
the alignment parameters from
flavor-changing $\tau$ decays, $Z\to b\bar b$,
$\epsilon_K$, and the radiative $b\to s\gamma$ decay.
By introducing the basis-independent Yukawa delay factor 
$\Delta_{H_1\bar f f}\equiv |\zeta_{f}|(1-g_{_{H_1VV}}^2)^{1/2}$,
we scrutinize the alignment of the Yukawa couplings of the 
lightest Higgs boson to the SM fermions.
\end{spacing}
\end{abstract}

\maketitle

\section{Introduction}
Since the discovery of the 125 GeV Higgs boson in 2012 at the
LHC \cite{Aad:2012tfa,Chatrchyan:2012ufa}, it has been inspected
very closely and extensively.
At the early stage, several model-independent studies
\cite{Carmi:2012yp,Azatov:2012bz,Espinosa:2012ir,
Klute:2012pu,Carmi:2012zd,Low:2012rj,Giardino:2012dp,Ellis:2012hz,Espinosa:2012im,
Carmi:2012in,Banerjee:2012xc,Bonnet:2012nm,Plehn:2012iz,Djouadi:2012rh,Dobrescu:2012td,
Cacciapaglia:2012wb,Belanger:2012gc,Moreau:2012da,Corbett:2012dm,Corbett:2012ja,
Masso:2012eq,Cheung:2013kla,Cheung:2014noa}
show that there were some rooms for it to be unlike the one predicted
in the Standard Model (SM) but, after combining  all the LHC Higgs data
at 7 and 8 TeV~\cite{Khachatryan:2016vau} and especially those
at 13 TeV~\cite{ATLAS:2018uso,Sirunyan:2018ouh,ATLAS:2018bsg,CMS:2018mmw,
ATLAS:2018gcr,CMS:2018xuk,Aaboud:2018zhk,CMS:2018lkl,Sirunyan:2018kst,ATLAS:2018lur,
CMS:2018nqp,Aaboud:2018urx,Aaboud:2017jvq,Aaboud:2017rss,Sirunyan:2018shy,
Sirunyan:2018ygk,Sirunyan:2018mvw,Sirunyan:2018koj,Aad:2019mbh},
it turns out that it is best described by the SM Higgs boson.
Specifically, the third-generation Yukawa couplings have been established. And
the most recent model-independent study~\cite{Cheung:2018ave} 
shows that
the $1\sigma$ error of the top-quark Yukawa coupling is
about 6\% while those of the bottom-quark and tau-lepton ones are
about 10\%. 
\footnote{
Throughout this work, we are using
the results presented in Ref.~\cite{Cheung:2018ave} which are
based on global fits of the Higgs boson couplings to all the LHC Higgs data
at 7 TeV, 8 TeV, and 13 TeV available up to the Summer 2018, corresponding
to integrated luminosities per experiment of approximately
5/fb at 7 TeV, 20/fb at 8 TeV and up to 80/fb at 13 TeV.
We note that there are more datasets at 13 TeV up to 139/fb and 137/fb 
collected with the ATLAS and CMS experiments, respectively, 
see Refs.~\cite{ATLAS:2020qdt,CMS:2020gsy}.
Though, without a combined ATLAS and CMS analysis, 
it is difficult to say conclusively
how much the full 13-TeV dataset improves the measurements of Higgs boson 
properties quantitatively, we observe that
the $1\sigma$ errors are reduced by the amount of about 30\% by comparing the 
results presented in Ref.~\cite{ATLAS:2020qdt} with those in 
Ref.~\cite{Aad:2019mbh} in which the dataset up to 80/fb is used.
}
In addition, the possibility of negative top-quark Yukawa coupling
has been completely ruled out and the bottom-quark Yukawa coupling
shows a preference of the positive sign
\footnote{
Precisely speaking, here the sign of the bottom-quark Yukawa coupling
is relative to the top-quark Yukawa coupling
configured through the $b$- and $t$-quark loop contributions to
the $Hgg$ vertex.
}
at about $1.5\sigma$ level.
For the tau-Yukawa coupling, the current data
still do not show any preference for its sign yet.
On the other hand, the coupling to a pair of massive vector bosons
is constrained to be consistent with the SM value
within about 5\% at $1\sigma$ level.

Even though we have not seen any direct hint or evidence of new physics beyond the SM (BSM),
we are eagerly anticipating it with various compelling motivations 
such as the tiny but non-vanishing neutrino
masses, matter dominance of our Universe and its evolution driven by dark energy and
dark matters, etc~\cite{Khlopov:2021xnw}.
In many BSM models, the Higgs sector is extended and it results in
existence of several neutral and charged Higgs bosons.
Their distinctive features depending on new theoretical frameworks
could be directly probed through their productions and decays at
future high-energy and high-precision experiments
\cite{Gunion:1989we,Gunion:1992hs,Carena:2002es,
Djouadi:2005gi,Djouadi:2005gj,Accomando:2006ga,Eriksson:2009ws,
Dittmaier:2011ti,Dittmaier:2012vm,Heinemeyer:2013tqa,deFlorian:2016spz,
Dawson:2013bba,Spira:1997dg,Spira:2016ztx,Dawson:2018dcd,Choi:2021nql}.

By the alignment of the Yukawa couplings in general 2HDMs
\cite{Lee:1973iz,Lee:1974jb,Peccei:1977hh,Fayet:1974fj,Inoue:1982ej,Flores:1982pr,
Gunion:1984yn,Botella:1994cs,Branco:1999fs,Branco:2011iw},
first of all, we imply that
the Yukawa matrices describing the couplings of the
two Higgs doublets to the SM fermions should be aligned
in the flavor space to avoid
the tree-level Higgs-mediated
flavor-changing neutral current (FCNC).
In 2HDMs, there are three neutral Higgs bosons and one of them
should be identified as the observed one at the LHC which weighs
125.5 GeV~\cite{Sirunyan:2020xwk}.
In this case, by the alignment of the Yukawa couplings, we also mean that
the couplings of this SM-like Higgs boson to the SM fermions
should be the same as those of the SM Higgs boson itself or
its couplings are strongly
constrained to be very SM-like by the current LHC data
as outlined above.
One of the popular ways to achieve this alignment
is to identify the lightest neutral Higgs boson as the 125.5 GeV one
and assume that all the other Higgs bosons are heavier or much heavier than
the lightest one~\cite{Haber:1989xc,Gunion:2002zf}.
But this decoupling scenario is not phenomenologically interesting and
another scenario is suggested in which all the couplings of the SM-like Higgs candidate
are (almost) aligned with those of the SM Higgs while the other Higgs bosons
are not so heavy~\cite{Craig:2013hca,Carena:2013ooa,Dev:2014yca,Bernon:2015qea}.

The alignment of Yukawa couplings are previously discussed and studied
\cite{Carena:2013ooa,Carena:2014nza}.
For some recent works, see, for example,
Refs.~\cite{Grzadkowski:2018ohf,Kanemura:2020ibp,Low:2020iua,Li:2020dbg}.
In this work, taking the framework of general 2HDMs,
we implement a comprehensive and detailed study of
the alignment of Yukawa couplings in the so-called Higgs basis
\cite{Donoghue:1978cj,Georgi:1978ri,Botella:1994cs,Branco:1999fs,
Davidson:2005cw,Haber:2006ue,Boto:2020wyf}
in which only the doublet containing the SM-like Higgs boson
develops the non-vanishing vacuum expectation value (vev) $v$.
For the alignment of the Yukawa matrices, we assume that the Yukawa matrices
are aligned in the flavor space
\cite{Manohar:2006ga,Pich:2009sp,Penuelas:2017ikk}
by introducing the
three alignment parameters $\zeta_f$ with $f=u,d,e$
for the couplings to the up-type quarks, the down-type quarks,
and the charged leptons, respectively.
Under this assumption, there are no
Higgs-mediated FCNC couplings at tree level and, at higher orders,
they are very suppressed
\cite{Pich:2009sp,Penuelas:2017ikk,Jung:2010ik,Braeuninger:2010td,
Bijnens:2011gd}.
And then,
we identify the lightest neutral Higgs boson as the 125.5 GeV one and
consider the alignment of its Yukawa couplings as the masses of the
heavier Higgs bosons increase or as the heavy Higgs bosons decouple.
We configure that the decoupling of the Yukawa couplings
of the lightest Higgs boson is delayed by the amount of
$\Delta_{H_1\bar f f}\equiv 
|\zeta_{f}|(1-g_{_{H_1VV}}^2)^{1/2}$ compared to
its  coupling to a pair of massive vector bosons, $g_{_{H_1VV}}$.
We observe that the Yukawa delay factor $\Delta_{H_1\bar f f}$
can be sizable even when $g_{_{H_1VV}} \sim 1$
if $|\zeta_f|$ is significantly larger than $1$.
We consider the upper limit on $|\zeta_{u}|$ 
from $Z\to b \bar b$ and $\epsilon_K$, 
and, for $|\zeta_{d}|$ and $|\zeta_{e}|$,
we demonstrate that they  are constrained to be small
by the precision LHC Higgs data unless
the so-called wrong-sign alignment of the Yukawa couplings
\cite{Gunion:2002zf,Ferreira:2014naa,Ferreira:2014dya,Biswas:2015zgk,
Coyle:2018ydo} occurs.
\footnote{In the wrong-sign alignment limit, the Yukawa couplings are
equal in strength but opposite in sign to the SM ones.}
Note that the Yukawa delay factor is basis-independent and 
can be used even when
some of the Higgs potential parameters and/or all of the
three alignment parameters are complex.

We emphasize that we are reconsidering the decoupling behavior of the 
Yukawa couplings in the light of the new basis-independent measure 
of the Yukawa delay factor $\Delta_{H_1\bar f f}$
taking the aligned 2HDM in the Higgs basis.
In the Higgs basis, contrasting to the relatively well-known $\Phi$ basis, it is
easier to understand the analytic structure of
intercorrelations among the model parameters.
On the other hand, in the aligned 2HDM,
there are three uncorrelated complex alignment parameters 
which provide further CP-violating 
sources in addition to those in the Higgs potential.
The aligned 2HDM accommodates the conventional four types of 
2HDMs as the limiting cases
when the alignment parameters are real and fully correlated.

This paper is organized as follows.
Section II is devoted to a brief review of the 2HDM Higgs potential,
the mixing among neutral Higgs bosons and their couplings  to the SM particles in the Higgs
basis.
In Section III, we elaborate on the constraints from the perturbative unitarity,
the Higgs potential bounded from below, and the electroweak precision observables
as well as the flavor constraints on the alignment parameters.
And we carry out numerical analysis of the constraints and
the alignment of Yukawa couplings in Section IV.
A brief summary and conclusions are made in Section V.

\section{Two Higgs Doublet Model in the Higgs Basis}
In this section, we study the two Higgs doublet model
taking the so-called Higgs basis
\cite{Donoghue:1978cj,Georgi:1978ri,
Botella:1994cs,Branco:1999fs,Davidson:2005cw,Haber:2006ue,Boto:2020wyf}.
We consider the general potential containing 3 dimensionful
quadratic and 7 dimensionless quartic parameters, of which four
parameters are complex.
We closely examine the relations among the potential parameters,
Higgs-boson masses, and the neutral Higgs-boson mixing so as to 
figure out the set of input parameters to be used in the next Section.
We further work out the Yukawa couplings in the Higgs basis together with
the interactions of the neutral and charged Higgs bosons with
massive gauge bosons.
\subsection{Higgs Potential}
The general 2HDM scalar potential containing two
complex SU(2)$_L$ doublets of $\Phi_1$ and $\Phi_2$ with the same
hypercharge $Y=1/2$ may be given by~\cite{Choi:2021nql}
\footnote{In contrast with the Higgs 
basis which has been taken for this work, we address it as the $\Phi$ basis.}
\begin{eqnarray}
\label{eq:VPhi}
V_\Phi &=&
\mu_1^2 (\Phi_1^{\dagger} \Phi_1)
+\mu_2^2 (\Phi_2^{\dagger} \Phi_2)
+m_{12}^2 (\Phi_1^{\dagger} \Phi_2)
+m_{12}^{*2}(\Phi_2^{\dagger} \Phi_1) \nonumber \\
&&+ \lambda_1 (\Phi_1^{\dagger} \Phi_1)^2 + \lambda_2
(\Phi_2^{\dagger} \Phi_2)^2 + \lambda_3 (\Phi_1^{\dagger}
\Phi_1)(\Phi_2^{\dagger} \Phi_2) + \lambda_4 (\Phi_1^{\dagger}
\Phi_2)(\Phi_2^{\dagger} \Phi_1) \nonumber \\
&&+ \lambda_5 (\Phi_1^{\dagger} \Phi_2)^2 +
\lambda_5^{*} (\Phi_2^{\dagger} \Phi_1)^2 + \lambda_6
(\Phi_1^{\dagger} \Phi_1) (\Phi_1^{\dagger} \Phi_2) + \lambda_6^{*}
(\Phi_1^{\dagger} \Phi_1)(\Phi_2^{\dagger} \Phi_1) \nonumber \\
&& + \lambda_7 (\Phi_2^{\dagger} \Phi_2) (\Phi_1^{\dagger} \Phi_2) +
\lambda_7^{*} (\Phi_2^{\dagger} \Phi_2) (\Phi_2^{\dagger} \Phi_1)\; ,
\end{eqnarray}
in terms of 2 real and 1 complex dimensionful quadratic
couplings and
4 real and 3 complex dimensionless quartic couplings.
Note that the ${\mathbf Z}_2$ symmetry under $\Phi_1 \to \pm\Phi_1$ and
$\Phi_2 \to \mp\Phi_2$ is hardly broken
by the non-vanishing quartic couplings $\lambda_6$ and $\lambda_7$
and, in this case, we have three rephasing-invariant CP-violating
phases in the potential.
With the general parameterization of two scalar doublets $\Phi_{1,2}$ 
as
\begin{equation}
\Phi_1=\left(\begin{array}{c}
\phi_1^+ \\ \frac{1}{\sqrt{2}}\,(v_1+\phi_1+ia_1)
\end{array}\right)\,; \ \ \
\Phi_2={\rm e}^{i\xi}\,\left(\begin{array}{c}
\phi_2^+ \\ \frac{1}{\sqrt{2}}\,(v_2+\phi_2+ia_2)
\end{array}\right)\,,
\label{eq:general_parameterization}
\end{equation}
and denoting $v_1=v \cos\beta=vc_\beta$ and $v_2=v \sin\beta=vs_\beta$
with $v=\sqrt{v_1^2+v_2^2}$,
one may remove $\mu_1^2$, $\mu_2^2$,
and $\imag(m_{12}^2{\rm e}^{i\xi})$ from the 2HDM potential using three tadpole
conditions:
\begin{eqnarray}
\label{eq:2hdmtadpole}
\mu_1^2 &=& - v^2\left[\lambda_1c_\beta^2+\frac{1}{2}\lambda_3s_\beta^2+
c_\beta s_\beta\real(\lambda_6{\rm e}^{i\xi})\right]+s_\beta^2 M_{H^\pm}^2\,,
\nonumber \\
\mu_2^2 &=& - v^2\left[\lambda_2s_\beta^2+\frac{1}{2}\lambda_3c_\beta^2+
c_\beta s_\beta\real(\lambda_7{\rm e}^{i\xi})\right]+c_\beta^2 M_{H^\pm}^2\,,
\nonumber \\
\imag(m_{12}^2{\rm e}^{i\xi}) &=& -
\frac{v^2}{2}\left[
2\ c_\beta s_\beta\imag(\lambda_5{\rm e}^{2i\xi})+
c_\beta^2\imag(\lambda_6{\rm e}^{i\xi})+
s_\beta^2\imag(\lambda_7{\rm e}^{i\xi})
\right]\,,
\label{eq:tadpole_condition_in_phi_basis}
\end{eqnarray}
with the square of the charged Higgs-boson mass
\begin{equation}
\label{eq:2hdm_mch2}
M_{H^\pm}^2=
-\frac{\real(m_{12}^2{\rm e}^{i\xi})}{c_\beta s_\beta}
-\frac{v^2}{2c_\beta s_\beta}\left[\lambda_4 c_\beta s_\beta+
2\ c_\beta s_\beta\real(\lambda_5{\rm e}^{2i\xi})+
c_\beta^2\real(\lambda_6{\rm e}^{i\xi})+
s_\beta^2\real(\lambda_7{\rm e}^{i\xi})
\right]\,.
\end{equation}
On the other hand, in the Higgs basis where 
only one doublet contains the non-vanishing vev $v$,
the general 2HDM scalar potential again contains
three (two real and one complex) massive parameters
and four real and three complex dimensionless quartic couplings
and it might take the same form as in the $\Phi$ basis:
\begin{eqnarray}
\label{eq:VHiggs}
V_{\cal H} &=&
Y_1 ({\cal H}_1^{\dagger} {\cal H}_1)
+Y_2 ({\cal H}_2^{\dagger} {\cal H}_2)
+Y_3 ({\cal H}_1^{\dagger} {\cal H}_2)
+Y_3^{*}({\cal H}_2^{\dagger} {\cal H}_1) \nonumber \\
&&+ Z_1 ({\cal H}_1^{\dagger} {\cal H}_1)^2 + Z_2
({\cal H}_2^{\dagger} {\cal H}_2)^2 + Z_3 ({\cal H}_1^{\dagger}
{\cal H}_1)({\cal H}_2^{\dagger} {\cal H}_2) + Z_4 ({\cal H}_1^{\dagger}
{\cal H}_2)({\cal H}_2^{\dagger} {\cal H}_1) \nonumber \\
&&+ Z_5 ({\cal H}_1^{\dagger} {\cal H}_2)^2 +
Z_5^{*} ({\cal H}_2^{\dagger} {\cal H}_1)^2 + Z_6
({\cal H}_1^{\dagger} {\cal H}_1) ({\cal H}_1^{\dagger} {\cal H}_2) + Z_6^{*}
({\cal H}_1^{\dagger} {\cal H}_1)({\cal H}_2^{\dagger} {\cal H}_1) \nonumber \\
&& + Z_7 ({\cal H}_2^{\dagger} {\cal H}_2) ({\cal H}_1^{\dagger} {\cal H}_2) +
Z_7^{*} ({\cal H}_2^{\dagger} {\cal H}_2) ({\cal H}_2^{\dagger} {\cal H}_1)\; ,
\end{eqnarray}
where the new complex SU(2)$_L$ doublets of ${\cal H}_1$ and ${\cal H}_2$
are given by the linear combinations of $\Phi_1$ and $\Phi_2$ as follows
\begin{eqnarray}
\label{eq:H12inHiggBasis}
{\cal H}_1&=&c_\beta\Phi_1+{\rm e}^{-i\xi}s_\beta\Phi_2=
\left(\begin{array}{c}
G^+ \\ \frac{1}{\sqrt{2}}\,(v+\varphi_1+iG^0)
\end{array}\right)\,; \nonumber \\[2mm]
{\cal H}_2&=&-s_\beta\Phi_1+{\rm e}^{-i\xi}c_\beta\Phi_2=
\left(\begin{array}{c}
H^+ \\ \frac{1}{\sqrt{2}}\,(\varphi_2+ia)
\end{array}\right)\,,
\end{eqnarray}
with the relations 
\begin{equation}
\varphi_1\equiv c_\beta\phi_1+s_\beta\phi_2\,, \ \ \
\varphi_2\equiv -s_\beta\phi_1+c_\beta\phi_2\,; \ \ \
a=-s_\beta a_1+ c_\beta a_2\,,
\end{equation}
in terms of $\phi_{1,2}$ and $a_{1,2}$ 
in Eq.~(\ref{eq:general_parameterization}).
Incidentally, we have that
$G^0=c_\beta a_1 +s_\beta a_2$,
$G^+=c_\beta \phi_1^+ +s_\beta \phi_2^+$, and
$H^+=-s_\beta \phi_1^+ +c_\beta \phi_2^+$.
Note that only the neutral component of
the ${\cal H}_1$ doublet develops
the non-vanishing vacuum expectation value $v$ and
it contains only one physical degree of freedom let alone
the Goldstone modes.
In the so-called decoupling limit, ${\cal H}_1$ take over
the role of the SM SU(2)$_L$ doublet and
the remaining three Higgs states are accommodated
only by the ${\cal H}_2$ doublet.
\footnote{
For a numerical study later, the notations of
$\varphi_1=h$, $\varphi_2=H$, and $a=A$ are taken
in the decoupling limit.}

The potential parameters $Y_{1,2,3}$ and $Z_{1-7}$ in the Higgs basis
could be related to those in the $\Phi$ basis through: 
\begin{eqnarray}
Y_1&=&
\mu_1^2 c_\beta^2 + \mu_2^2 s_\beta^2 +
\real(m_{12}^2{\rm e}^{i\xi})s_{2\beta}\,, \nonumber \\[2mm]
Y_2&=&
\mu_1^2 s_\beta^2 + \mu_2^2 c_\beta^2 -
\real(m_{12}^2{\rm e}^{i\xi}) s_{2\beta}\,, \nonumber \\[2mm]
Y_3&=&
-(\mu_1^2-\mu_2^2) c_\beta s_\beta +
\real(m_{12}^2{\rm e}^{i\xi})c_{2\beta} + i\,
\imag(m_{12}^2{\rm e}^{i\xi}) \,,
\end{eqnarray}
for two real and one complex dimensionful parameters
and~\footnote{
We find that our results are consistent with those
presented in, for example, Ref.~\cite{Kanemura:2020ibp}.}
\begin{eqnarray}
Z_1&=&\lambda_1 c_\beta^4 +\lambda_2 s_\beta^4 +2\lambda_{345}c_\beta^2 s_\beta^2
+\left[\real(\lambda_6{\rm e}^{i\xi})c_\beta^2
 +\real(\lambda_7{\rm e}^{i\xi})s_\beta^2\right]s_{2\beta}\,, \nonumber \\[2mm]
Z_2&=&\lambda_1 s_\beta^4 +\lambda_2 c_\beta^4 +2\lambda_{345}c_\beta^2 s_\beta^2
-\left[\real(\lambda_6{\rm e}^{i\xi})s_\beta^2
 +\real(\lambda_7{\rm e}^{i\xi})c_\beta^2\right] s_{2\beta}\,, \nonumber \\[2mm]
Z_3&=&\lambda_3 + 2(\lambda_1+\lambda_2-2\lambda_{345}) c_\beta^2 s_\beta^2
-\left[\real(\lambda_6{\rm e}^{i\xi})-\real(\lambda_7{\rm e}^{i\xi})\right]
c_{2\beta} s_{2\beta}\,, \nonumber \\[2mm]
Z_4&=&\lambda_4 + 2(\lambda_1+\lambda_2-2\lambda_{345}) c_\beta^2 s_\beta^2
-\left[\real(\lambda_6{\rm e}^{i\xi})-\real(\lambda_7{\rm e}^{i\xi})\right]
c_{2\beta}s_{2\beta}\,, \nonumber \\[2mm]
Z_5&=&(\lambda_1+\lambda_2-2\lambda_{345}) c_\beta^2 s_\beta^2
+\real(\lambda_5{\rm e}^{2i\xi})
-\left[\real(\lambda_6{\rm e}^{i\xi})-\real(\lambda_7{\rm e}^{i\xi})\right]
c_{2\beta}c_\beta s_\beta \nonumber \\
&& +\,i\left[\imag(\lambda_5{\rm e}^{2i\xi})c_{2\beta}
-\imag(\lambda_6{\rm e}^{i\xi})c_\beta s_\beta
+\imag(\lambda_7{\rm e}^{i\xi})c_\beta s_\beta \right]\,, \nonumber \\[2mm]
Z_6&=&(-\lambda_1 c_\beta^2 +\lambda_2 s_\beta^2)s_{2\beta}
+2\lambda_{345}c_{2\beta} c_\beta s_\beta
+\real(\lambda_6{\rm e}^{i\xi})(c_\beta^2 - 3s_\beta^2)c_\beta^2
+\real(\lambda_7{\rm e}^{i\xi})(3c_\beta^2 - s_\beta^2)s_\beta^2
\nonumber \\ &&
\,+\,i\left[\imag(\lambda_5{\rm e}^{2i\xi})s_{2\beta}
+\imag(\lambda_6{\rm e}^{i\xi})c_\beta^2
+\imag(\lambda_7{\rm e}^{i\xi})s_\beta^2 \right]\,, \nonumber \\[2mm]
Z_7&=&(-\lambda_1 s_\beta^2 +\lambda_2 c_\beta^2) s_{2\beta}
-2\lambda_{345}c_{2\beta} c_\beta s_\beta
+\real(\lambda_6{\rm e}^{i\xi})(3c_\beta^2 - s_\beta^2)s_\beta^2
+\real(\lambda_7{\rm e}^{i\xi})(c_\beta^2 - 3s_\beta^2)c_\beta^2
\nonumber \\ &&
\,+\,i\left[-\imag(\lambda_5{\rm e}^{2i\xi}) s_{2\beta}
+\imag(\lambda_6{\rm e}^{i\xi})s_\beta^2
+\imag(\lambda_7{\rm e}^{i\xi})c_\beta^2 \right]\,,
\end{eqnarray}
for four real and three complex dimensionless parameters
with $\lambda_{345}\equiv (\lambda_3+\lambda_4)/2+\real(\lambda_5{\rm e}^{2i\xi})$.
We note that $Z_1\leftrightarrow Z_2$,
$Z_3\leftrightarrow Z_4$, $Z_6\leftrightarrow Z_7$
and $Z_5$ is invariant under the exchanges
$c_\beta \leftrightarrow s_\beta$,
$\lambda_3 \leftrightarrow \lambda_4$,
$(\lambda_5{\rm e}^{2i\xi}) \leftrightarrow (\lambda_5{\rm e}^{2i\xi})^*$,
$(\lambda_{6,7}{\rm e}^{i\xi}) \leftrightarrow -(\lambda_{6,7}{\rm e}^{i\xi})^*$.
The tadpole conditions in the Higgs basis, which are 
much simpler than those in the $\Phi$ basis as shown
in Eq.~(\ref{eq:tadpole_condition_in_phi_basis}), are
\begin{equation}
\label{eq:higgsbasistadpole}
Y_1 \ + \ Z_1 v^2\ = 0 \,; \ \ \
Y_3 \ + \ \frac{1}{2}Z_6 v^2\ = 0 \,,
\end{equation}
where the first condition comes from
$\langle\frac{\partial V_{\cal H}}{\partial\varphi_1}\rangle=0$
and the second one from
$\langle\frac{\partial V_{\cal H}}{\partial\varphi_2}\rangle=0$
and $\langle\frac{\partial V_{\cal H}}{\partial a}\rangle=0$.
Note that the second condition relates the two complex parameters
of $Y_3$ and $Z_6$.

\subsection{Masses, Mixing, and Potential Parameters in the Higgs Basis}
In the Higgs basis, the 2HDM Higgs potential includes the mass
terms which can be cast into the form consisting of two parts
\begin{equation}
V_{{\cal H}\,, {\rm mass}}=
M_{H^\pm}^2 H^+ H^- \ + \ \frac{1}{2}
(\varphi_1 \ \varphi_2  \ a)\,{\cal M}^2_0\,
\left(\begin{array}{c}
\varphi_1 \\ \varphi_2  \\ a \end{array}\right)\,,
\end{equation}
in terms of the charged Higgs bosons $H^\pm$, two
neutral scalars $\varphi_{1,2}$, and one neutral pseudoscalar $a$.
The charged Higgs boson mass is given by
\begin{equation}
M_{H^\pm}^2= Y_2 +\frac{1}{2} Z_3 v^2\,,
\end{equation}
while the $3\times 3$ mass-squared matrix of the neutral Higgs
bosons ${\cal M}_0^2$ takes the form
\begin{equation}
{\cal M}^2_0 = M_A^2 \ {\rm diag}(0,1,1) \ + \ {\cal M}^2_Z\,,
\end{equation}
where
$M_A^2=M_{H^\pm}^2+ \left[\frac{1}{2}Z_4 -\real(Z_5)\right]v^2$ and
the $3\times 3$ real  and symmetric mass-squared matrix ${\cal M}^2_Z$
is given by
\begin{eqnarray}
\frac{{\cal M}^2_Z}{v^2} &=& \left(\begin{array}{ccc}
2 Z_1 & \real(Z_6) & -\imag(Z_6) \\
\real(Z_6) & 2\real(Z_5) & -\imag(Z_5) \\
-\imag(Z_6) & -\imag(Z_5) & 0
\end{array}\right)\,.
\end{eqnarray}
Note that the quartic couplings $Z_2$ and $Z_7$ have nothing to
do with the masses of Higgs bosons and the mixing of the neutral ones.
They can be probed only through the cubic and quartic Higgs self-couplings,
see Eq.~(\ref{eq:VHiggs})
while noting that only the ${\cal H}_1$ doublet contains 
the vev $v\simeq 246$ GeV.
We further note that
$\varphi_1$ decouples from the mixing with
the other two neutral states of $\varphi_2$ and $a$ in the $Z_6=0$ limit,
and its mass squared is simply given by $2 Z_1 v^2$ which
gives $Z_1\simeq 0.13\,(M_{H_1}/125.5\,{\rm GeV})^2$.
And, in this decoupling limit of $Z_6\to 0$, the CP-violating mixing between
the two states of $\varphi_2$ and $a$ is dictated only by
$\imag(Z_5)$.

Once the $3\times 3$ real and symmetric mass-squared
matrix ${\cal M}_0^2$ is given,
the orthogonal $3\times 3$ mixing matrix $O$ is defined through
\footnote{Note that we reserve the notations of
$H_{i=1,2,3}$ for the mass eigenstates of
three neutral Higgs bosons taking account of CP-violating mixing
in the neutral Higgs-boson sector when $\imag(Z_{5,6})\neq 0$.
In general, the neutral Higgs bosons
do not carry definite CP parities and they become mixtures of CP-even and CP-odd
states.}
\begin{eqnarray}
(\varphi_1,\varphi_2,a)^T_\alpha&=&O_{\alpha i} (H_1,H_2,H_3)^T_i\,,
\end{eqnarray}
such that $O^T {\cal M}_0^2 O={\rm diag}(M_{H_1}^2,M_{H_2}^2,M_{H_3}^2)$
with the increasing ordering of $M_{H_1}\leq M_{H_2}\leq M_{H_3}$,
if necessary.
Note that the mass-squared matrix
${\cal M}_0^2$ involves only the four (two real and two complex)
quartic couplings $\{Z_1,Z_4,Z_5,Z_6\}$
once $v$ and $M_{H^\pm}$ are given.
And then, using the matrix relation
$O^T {\cal M}_0^2 O={\rm diag}(M_{H_1}^2,M_{H_2}^2,M_{H_3}^2)$,
one may find the following expressions for
the quartic couplings of $\{Z_1,Z_4,Z_5,Z_6\}$
in terms of the three masses of neutral Higgs bosons and
the components of the $3\times 3$ orthogonal mixing matrix $O$:
\footnote{The $3\times 3$ orthogonal mixing matrix $O$
contains three independent degrees of freedom
represented by the three rotation angles.}
\begin{eqnarray}
\label{eq:Z1456}
Z_1 &=& \frac{1}{2 v^2} \left(
M_{H_1}^2 O_{\varphi_11}^2 + M_{H_2}^2 O_{\varphi_12}^2 +
M_{H_3}^2 O_{\varphi_13}^2 \right) \,, \nonumber \\[2mm]
Z_4 &=& \frac{1}{v^2}
\left[M_{H_1}^2 (O_{\varphi_21}^2 + O_{a1}^2) +
M_{H_2}^2 (O_{\varphi_22}^2 + O_{a2}^2)+
M_{H_3}^2 (O_{\varphi_23}^2 + O_{a3}^2) - 2 M_{H^\pm}^2
\right] \,, \nonumber \\[2mm]
Z_5 &=& \frac{1}{2 v^2}
\left[M_{H_1}^2 (O_{\varphi_21}^2 - O_{a1}^2) +
M_{H_2}^2 (O_{\varphi_22}^2 - O_{a2}^2)+
M_{H_3}^2 (O_{\varphi_23}^2 - O_{a3}^2) \right]  \nonumber \\
&& -\frac{i}{v^2} \left(
M_{H_1}^2 O_{\varphi_21}O_{a1} + M_{H_2}^2 O_{\varphi_22}O_{a2}
+ M_{H_3}^2 O_{\varphi_23}O_{a3} \right) \,, \nonumber \\[2mm]
Z_6 &=& \frac{1}{v^2} \left(
M_{H_1}^2 O_{\varphi_11}O_{\varphi_21} + M_{H_2}^2 O_{\varphi_12}O_{\varphi_22} +
M_{H_3}^2 O_{\varphi_13}O_{\varphi_23} \right) \nonumber \\
&& -\frac{i}{v^2} \left(
M_{H_1}^2 O_{\varphi_11}O_{a1} + M_{H_2}^2 O_{\varphi_12}O_{a2} +
M_{H_3}^2 O_{\varphi_13}O_{a3} \right) \,,
\end{eqnarray}
for given $v$ and $M_{H^\pm}$.

Now we are ready to consider the input parameters 
for 2HDM in the Higgs basis. First of all,
the input parameters for the Higgs potential Eq.~(\ref{eq:VHiggs}) are
\begin{equation}
\left\{Y_1,Y_2,Y_3;
Z_1,Z_2,Z_3,Z_4,Z_5,Z_6,Z_7\right\}\,.
\end{equation}
Using the tadpole conditions in 
Eq.~(\ref{eq:higgsbasistadpole}), the dimensionful parameters
$Y_1$ and $Y_3$ can be removed from
the set in favor of $v$ and observing that
the quartic couplings $Z_2$ and $Z_7$ do not contribute to
the mass terms, one may consider the following set
of input parameters
\begin{equation}
\label{eq:y2zi}
\left\{v,Y_2;M_{H^\pm},
Z_1,Z_4,Z_5,Z_6;Z_2,Z_7\right\}\,,
\end{equation}
where we trade the quartic coupling $Z_3$ with
the charged Higgs mass $M_{H^\pm}$ using the relation
$Z_3 =2\left(M_{H^\pm}^2 - Y_2\right)/v^2$ with $Y_2$ given.
Further using $M_{H_i}$ and $O$ instead of
$\{Z_1,Z_4,Z_5,Z_6\}$, we end up with the following
set of input parameters:
\begin{equation}
{\cal I}=
\left\{v,Y_2;M_{H^\pm},M_{H_1},M_{H_2},M_{H_3},\{O_{3\times 3}\};Z_2,Z_7\right\}\,,
\end{equation}
which contains 12 real degrees of freedom.
If desirable, one may remove the 
unphysical massive parameter $Y_2$ in favor of the dimensionless
quartic coupling $Z_3$ by having an alternative set
\begin{equation}
{\cal I}^\prime=
\left\{v;M_{H^\pm},M_{H_1},M_{H_2},M_{H_3},\{O_{3\times 3}\};Z_3;Z_2,Z_7\right\}\,,
\end{equation}
consisting of 12 real parameters as well.

For example, in the CP-conserving (CPC) case with $\imag{Z_5}=\imag{Z_6}=0$,
one may denote
the masses of the three neutral Higgs bosons by
$M_h$, $M_H$, and $M_A$
or $O^T {\cal M}_0^2 O={\rm diag}(M_{h}^2,M_{H}^2,M_{A}^2)$.
Note that $M_h^2=2Z_1v^2$ is
for the SM Higgs boson in the decoupling limit of $Z_6\to 0$.
The mixing matrix $O$ can be parameterized as
\begin{equation}
O_{\rm CPC} \ = \ \left(\begin{array}{ccc}
c_\gamma & s_\gamma & 0 \\
-s_\gamma & c_\gamma & 0 \\
0 & 0 & 1 \end{array}\right)\,,
\end{equation}
introducing the mixing angle $\gamma$ between the
two CP-even states $\varphi_1$ and $\varphi_2$.
In this CP-conserving case,
the relations Eq.~(\ref{eq:Z1456}) simplify into
\begin{eqnarray}
\label{eq:Z1456_cpc}
Z_1 &=& \frac{1}{2v^2}\left(c_\gamma^2 M_h^2 + s_\gamma^2 M_H^2 \right)\,, \ \ \
\hspace{1.1cm}
Z_4  =  \frac{1}{v^2}\left( s_\gamma^2 M_h^2 + c_\gamma^2 M_H^2
+M_A^2 -2M_{H^\pm}^2 \right)\,,\nonumber \\[2mm]
Z_5 &=& \frac{1}{2v^2}\left(s_\gamma^2 M_h^2
+c_\gamma^2 M_H^2 -M_A^2 \right)\,, \ \ \
Z_6  =  \frac{1}{v^2}\left(-M_h^2 + M_H^2 \right)c_\gamma s_\gamma\,.
\end{eqnarray}
We observe that, in the decoupling limit of $\sin\gamma=0$,
$Z_1=M_h^2/2v^2$ and $Z_6=0$,
and $Z_4$ and $Z_5$ are determined by the mass differences of
$M_H^2+M_A^2-2M_{H^\pm}^2$ and $M_H^2-M_A^2$, respectively.
Finally, for the study of the CPC case,
one may choose one of the following two equivalent sets:
\begin{eqnarray}
\label{eq:input_cpc}
{\cal I}_{\rm CPC}&=&
\left\{v,Y_2;M_{H^\pm},M_h,M_H,M_A,\gamma ;Z_2,Z_7\right\}\,,
\nonumber \\[2mm]
{\cal I}_{\rm CPC}^\prime&=&
\left\{v;M_{H^\pm},M_h,M_H,M_A,\gamma ;Z_3;Z_2,Z_7\right\}\,,
\end{eqnarray}
each of which contains 9 real degrees of freedom,
and the convention of $|\gamma|\leq \pi/2$
without loss of generality resulting in
$c_\gamma \geq 0$ and ${\rm sign}(s_\gamma)={\rm sign}(Z_6)$
if $M_H>M_h$ GeV.

In the presence of non-vanishing
$\imag{Z_5}$ and/or $\imag{Z_6}$, the mixing between the two CP-even states
$\varphi_{1,2}$ and the CP-odd one $a$ arises leading to 
CP violation in the neutral Higgs sector.
By introducing a rotation
${\cal H}_2 \to  {\rm e}^{i\zeta}\,{\cal H}_2$,
\footnote{Or, equivalently, ${\cal H}_1^\dagger{\cal H}_2 \to
{\rm e}^{i\zeta}\,{\cal H}_1^\dagger{\cal H}_2$.}
we note that the Higgs potential given by Eq.~(\ref{eq:VHiggs})
is invariant under the following phase rotations:
\begin{eqnarray}
\label{eq:rephaseZ567}
{\cal H}_2 \to  {\rm e}^{+i\zeta}\,{\cal H}_2\,; \ \
Y_3 \to Y_3\,{\rm e}^{-i\zeta}\,, \ \
Z_5 \to Z_5\,{\rm e}^{-2i\zeta}\,, \,
Z_6 \to Z_6\,{\rm e}^{-i\zeta}\,, \,
Z_7 \to Z_7\,{\rm e}^{-i\zeta}\,.
\end{eqnarray}
Considering the tadpole conditions Eq.~(\ref{eq:higgsbasistadpole}),
this might imply that 
one of the CP phases of $\imag(Z_5)$, $\imag(Z_6)$, and $\imag(Z_7)$
can be rotated away by rephasing the Higgs fields ${\cal H}_2$.
By keeping $\imag(Z_7)$ as an independent input
and taking either $\imag(Z_5)=0$ or $\imag(Z_6)=0$, one may
use the following set of input parameters:
\footnote{In this CP-violating (CPV) case, we parameterize
the mixing matrix $O$ by introducing the three mixing angles
of $\gamma$, $\eta$, and $\omega$ as explicitly 
shown in Eq.~(\ref{eq:omix}).}
\begin{equation}
\label{eq:ICPV0}
{\cal I}_{\rm CPV}=
\left\{v,Y_2;M_{H^\pm},M_{H_1},M_{H_2},M_{H_3},\gamma,
\{\omega ~{\rm or}~\eta\};Z_2,
\real(Z_7),\imag(Z_7)\right\}\,,
\end{equation}
which contains 11 real degrees of freedom.
In this case, the mixing angle $\eta$ ($\omega$) can be fixed by solving
$\imag(Z_5)=0$ or $\imag(Z_6)=0$ when $M_{H_{1,2,3}}$ and $\omega$ ($\eta$)
are given.
More explicitly, using the relations in Eq.~(\ref{eq:Z1456}), we have
\begin{eqnarray}
\imag(Z_5)&=&
\left[ \frac{M_{H_3}^2 c_\omega^2 +M_{H_2}^2 s_\omega^2 -M_{H_1}^2}{v^2}\,
s_\gamma s_\eta 
- \frac{M_{H_3}^2-M_{H_2}^2}{v^2}\, 
c_\gamma c_\omega s_{\omega}\,\right]\, c_\eta \nonumber \\[2mm]
&=&
\left[ \frac{M_{H_3}^2+ M_{H_2}^2 -2 M_{H_1}^2}{2v^2}\, s_\gamma s_\eta 
+ \frac{M_{H_3}^2-M_{H_2}^2}{2v^2}\, s_\gamma s_\eta c_{2\omega}
- \frac{M_{H_3}^2-M_{H_2}^2}{2v^2}\, c_\gamma s_{2\omega}\,\right]\, c_\eta 
\nonumber \\[3mm]
\imag(Z_6)&=& 
-\left[\frac{M_{H_3}^2 c_\omega^2 +M_{H_2}^2 s_\omega^2 -M_{H_1}^2}{v^2}\,
c_\gamma s_\eta 
+ \frac{M_{H_3}^2-M_{H_2}^2}{v^2}\, 
s_\gamma c_\omega s_{\omega}\,\right]\, c_\eta \nonumber \\[2mm]
&=&
-\left[ \frac{M_{H_3}^2+ M_{H_2}^2 -2 M_{H_1}^2}{2v^2}\, c_\gamma s_\eta 
+ \frac{M_{H_3}^2-M_{H_2}^2}{2v^2}\, c_\gamma s_\eta c_{2\omega}
+ \frac{M_{H_3}^2-M_{H_2}^2}{2v^2}\, s_\gamma s_{2\omega}\,\right]\, c_\eta \,,
\end{eqnarray}
parameterizing the mixing matrix $O$ as follow:
\begin{eqnarray}
\label{eq:omix}
O_{\rm CPV} = O_\gamma O_\eta O_\omega &\equiv &
\left( \begin{array}{ccc}
  c_\gamma  &   s_\gamma   &  0   \\
  -s_\gamma  &   c_\gamma  &  0   \\
  0         &      0       &  1   \\
  \end{array} \right)
\left( \begin{array}{ccc}
    c_\eta               &      0             &   s_\eta  \\
    0                    &      1             &   0   \\
   -s_\eta               &      0             &   c_\eta   \\
  \end{array} \right)
\left( \begin{array}{ccc}
    1               &      0       &   0   \\
    0               &   c_\omega   &   s_\omega   \\
    0               &  -s_\omega   &   c_\omega   \\
  \end{array} \right)
\nonumber \\[3mm] &=&
\left( \begin{array}{ccc}
    c_\gamma c_\eta  &  s_\gamma c_\omega - c_\gamma s_\eta s_\omega   &
    s_\gamma s_\omega + c_\gamma s_\eta c_\omega     \\
    -s_\gamma c_\eta   &  c_\gamma c_\omega + s_\gamma s_\eta s_\omega   &
    c_\gamma s_\omega - s_\gamma s_\eta c_\omega     \\
    -s_\eta  &  -c_\eta s_\omega   &    c_\eta c_\omega  \\
  \end{array} \right)\,.
\end{eqnarray}
Assuming $c_\eta \neq$ 0 and, for example, taking 
$\gamma$ and $\omega$ as the input mixing angles, 
the remaining mixing angle $\eta$ is determined by
\begin{equation}
\left.s_\eta\right|_{\imag{Z_5}=0}  
=  \frac{(M_{H_3}^2-M_{H_2}^2)c_\gamma c_\omega s_\omega}
{(M_{H_3}^2 c_\omega^2 +M_{H_2}^2 s_\omega^2 -M_{H_1}^2) s_\gamma}
\end{equation}
imposing $\imag{Z_5}=0$.
If $\imag{Z_6}=0$ is imposed instead, $\eta$ is determined by
\begin{equation}
\left.s_\eta\right|_{\imag{Z_6}=0}  
= -\frac{(M_{H_3}^2-M_{H_2}^2)s_\gamma c_\omega s_\omega}
{(M_{H_3}^2 c_\omega^2 +M_{H_2}^2 s_\omega^2 -M_{H_1}^2) c_\gamma}\,.
\end{equation}
Of course, using $Z_3$ instead of $Y_2$,
one may use the alternative set
\begin{equation}
\label{eq:ICPV1}
{\cal I}^\prime_{\rm CPV}=
\left\{v;M_{H^\pm},M_{H_1},M_{H_2},M_{H_3},\gamma,
\{\omega ~{\rm or}~\eta\};Z_3;Z_2,
\real(Z_7),\imag(Z_7)\right\}\,.
\end{equation}
Incidentally, one may choose the basis in which
$\imag(Z_7)=0$ by taking the following set of input
parameters:
\begin{equation}
\label{eq:ICPV2}
{\cal I}^{\prime\prime}_{\rm CPV}=
\left\{v;M_{H^\pm},M_{H_1},M_{H_2},M_{H_3},\gamma,
\eta,\omega;Z_3;Z_2,Z_7\right\}\,,
\end{equation}
where all the three mixing angles are independent from one another
and $Z_7$ is real.

In passing, we note that, in the limit of $c_\gamma = 1$ and $s_\gamma=0$,
the mixing matrix takes the simpler form
\begin{eqnarray}
\label{eq:omix_etaomega}
\left.O_{\rm CPV}\right|_{\sin\gamma=0} = O_\eta O_\omega  =
\left( \begin{array}{ccc}
    c_\eta  &   -  s_\eta s_\omega   & s_\eta c_\omega     \\
    0   &  c_\omega   &  s_\omega      \\
    -s_\eta  &  -c_\eta s_\omega   &    c_\eta c_\omega  \\
  \end{array} \right)\,.
\end{eqnarray}
When $c_\eta\simeq 1-\eta^2/2$  and $s_\eta \simeq \eta$,
the lightest $H_1$ is SM like and the heavier ones $H_{2,3}$
are mostly arbitrary mixtures of $\varphi_2$ and $a$.
On the other hand, when $c_\eta\simeq |\eta| $  and $|s_\eta| \simeq 1-\eta^2/2$,
the lightest $H_1$ is mostly CP odd ($H_1\sim a$) and
$H_2$ ($H_3$) is SM like when
$|s_\omega| \simeq 1$ ($|c_\omega| \simeq 1$).

\subsection{Yukawa Couplings in Higgs basis}
In the 2HDM, the Yukawa couplings might be given by~\cite{Pich:2009sp}
\begin{equation}
\label{eq:yukawa0}
-{\cal L}_Y=\sum_{k=1,2}\,
\overline{Q_L^0}\,{\bf y}^u_k\,\widetilde{\cal H}_k\,u_R^0 \ + \
\overline{Q_L^0}\,{\bf y}^d_k\,{\cal H}_k\,d_R^0 \ + \
\overline{L_L^0}\,{\bf y}^e_k\,{\cal H}_k\,e_R^0 \ + \ {\rm h.c.}
\end{equation}
in terms of the six $3\times 3$ Yukawa matrices  ${\bf y}_{1,2}^{u,d,e}$
with the electroweak eigenstates
$Q_L^0=(u_L^0\,, d_L^0)^T$, $L_L^0=(\nu_L^0\,, e_L^0)^T$, 
$u_R^0$, $d_R^0$, and $e_R^0$.
The two Higgs doublets ${\cal H}_{1,2}$ in the Higgs basis are given by
Eq.~(\ref{eq:H12inHiggBasis}):
\begin{eqnarray}
{\cal H}_1 = \left(G^+\,,\,\frac{1}{\sqrt{2}}(v+\varphi_1+i G^0)\right)^T\,, \ \ \
{\cal H}_2 = \left(H^+\,,\,\frac{1}{\sqrt{2}}(\varphi_2+i a)\right)^T\,,
\end{eqnarray}
and their SU(2)-conjugated doublets by
\begin{eqnarray}
\widetilde{\cal H}_1 = i\tau_2 {\cal H}_1^* =
\left(\frac{1}{\sqrt{2}}(v+\varphi_1-i G^0)\,,\,-G^-\right)^T\,, \ \ \
\widetilde{\cal H}_2 = i\tau_2 {\cal H}_2^* =
\left(\frac{1}{\sqrt{2}}(\varphi_2-i a)\,,\, -H^-\right)^T\,.
\end{eqnarray}
The Yukawa interactions include the following mass terms
\begin{equation}
-{\cal L}_{Y,\,{\rm mass}}=\frac{v}{\sqrt{2}}\left(
\overline{u_L^0}\,{\bf y}^u_1\,u_R^0 \ + \
\overline{d_L^0}\,{\bf y}^d_1\,d_R^0 \ + \
\overline{e_L^0}\,{\bf y}^e_1\,e_R^0 \ + \ {\rm h.c.}\right)\,,
\end{equation}
which involve only the Yukawa matrices of ${\bf y}^{u,d,e}_1$.
Therefore, introducing two unitary matrices 
relating the left/right-handed electroweak eigenstates 
$f_{L,R}^0$ to the left/right-handed mass eigenstates
$f_{L,R}$ with
$f=u,d,e$ as follows
\begin{eqnarray}
u_L^0  &=& {\cal U}_{u_L}\,u_L\,, \ \ \
d_L^0   =  {\cal U}_{d_L}\,d_L\,, \ \ \
e_L^0   =  {\cal U}_{e_L}\,e_L\,; \nonumber \\
u_R^0  &=& {\cal U}_{u_R}\,u_R\,, \ \ \
d_R^0   =  {\cal U}_{d_R}\,d_R\,, \ \ \
e_R^0   =  {\cal U}_{e_R}\,e_R\,,
\end{eqnarray}
we have, for the mass terms,
\begin{equation}
-{\cal L}_{Y,\,{\rm mass}}=
\overline{u_L}\,{\bf M}_u\,u_R \ + \
\overline{d_L}\,{\bf M}_d\,d_R \ + \
\overline{e_L}\,{\bf M}_e\,e_R \ + \ {\rm h.c.}\,,
\end{equation}
where the three diagonal matrices are
\begin{eqnarray}
{\bf M}_u &=& \frac{v}{\sqrt{2}}~{\cal U}_{u_L}^\dagger\, {\bf y}^u_1\, {\cal U}_{u_R}
={\rm diag}(m_u,m_c,m_t)\,, \nonumber \\
{\bf M}_d &=& \frac{v}{\sqrt{2}}~{\cal U}_{d_L}^\dagger\, {\bf y}^d_1\, {\cal U}_{d_R}
={\rm diag}(m_d,m_s,m_b)\,, \nonumber \\
{\bf M}_e &=& \frac{v}{\sqrt{2}}~{\cal U}_{e_L}^\dagger\, {\bf y}^e_1\, {\cal U}_{e_R}
={\rm diag}(m_e,m_\mu,m_\tau)\,,
\end{eqnarray}
in terms of the six quark and three charged-lepton masses.
We note that ${\cal U}_{u_L}^\dagger {\cal U}_{d_L}=V_{\rm CKM}\equiv V$
is nothing but the CKM matrix and, by the use of it, the SU(2)$_L$ quark doublets
in the electroweak basis can be related to
those in the mass basis in the following two ways:
\begin{equation}
Q_L^0\ =\ {\cal U}_{u_L}\left(
\begin{array}{c} u_L \\ V d_L\end{array}\right) \ \ {\rm or} \ \
Q_L^0\ = \ {\cal U}_{d_L}\left(
\begin{array}{c} V^\dagger u_L \\ d_L\end{array}\right)\,.
\end{equation}
The first relation is used for the Yukawa interactions
with the right-handed up-type quarks and the second one for those
with the right-handed down-type quarks.
Incidentally, we also have
\begin{equation}
L_L^0={\cal U}_{e_L}\left(
\begin{array}{c} \nu_L \\ e_L\end{array}\right)
\end{equation}
by defining $\nu_L\equiv {\cal U}_{e_L}^\dagger \nu_L^0$
with no physical effects in the case with vanishing
neutrino masses.

Collecting all the parameterizations, unitary rotations, 
and re-parameterizations, the couplings
of the neutral Higgs bosons to two fermions are given by
\begin{eqnarray}
-{\cal L}_{H\bar f f} &=&
\frac{1}{v}\left[\overline u\,{\bf M}_u\,u\right]\varphi_1  +
\left[\overline u\left({\bf h}_u^H +{\bf h}_u^A \gamma_5\right)u\right]\varphi_2 +
\left[\overline u\left(-i{\bf h}_u^A -i {\bf h}_u^H \gamma_5\right)u\right] a
\nonumber \\[2mm] &+&
\frac{1}{v}\left[\overline d\,{\bf M}_d\,d\right]\varphi_1  +
\left[\overline d\left({\bf h}_d^H +{\bf h}_d^A \gamma_5\right)d\right]\varphi_2 +
\left[\overline d\left(i{\bf h}_d^A +i {\bf h}_d^H \gamma_5\right)d\right] a
\nonumber \\[2mm] &+&
\frac{1}{v}\left[\overline e\,{\bf M}_e\,e\right]\varphi_1  +
\left[\overline e\left({\bf h}_e^H +{\bf h}_e^A \gamma_5\right)e\right]\varphi_2 +
\left[\overline e\left(i{\bf h}_e^A +i {\bf h}_e^H \gamma_5\right)e\right] a
\end{eqnarray}
where three Hermitian and three anti-Hermitian 
Yukawa coupling matrices are
\begin{equation}
{\bf h}_f^H  \equiv \frac{{\bf h}_f+{\bf h}_f^\dagger}{2}\,, \ \ \
{\bf h}_f^A \equiv \frac{{\bf h}_f-{\bf h}_f^\dagger}{2}\,,
\end{equation}
with ${\bf h}_{f=u,d,e}$ given in terms of
the $3\times 3$ Yukawa matrix ${\bf y}_2^f$
and two unitary matrices as
\begin{equation}
{\bf h}_f \equiv
\frac{1}{\sqrt{2}}~{\cal U}_{f_L}^\dagger\, {\bf y}_2^f\, {\cal U}_{f_R} \,.
\end{equation}
We observe that the couplings of the $\varphi_1$ field
are diagonal in the flavor space and their sizes are directly 
proportional to the masses of the fermions to which it couples.
In contrast, those of the $\varphi_2$ and $a$ fields 
are not diagonal in the flavor space leading to the
tree-level Higgs-mediated FCNC
and their magnitudes are arbitrary in principle.

To avoid the tree-level FCNC, the matrices ${\bf h}_{f=u,d,e}$ 
are desired to be diagonal which can be achieved by requiring~\cite{Pich:2009sp}
\footnote{Under this requirement, the Yukawa matrix ${\bf h}_f$
for the Higgs field ${\cal H}_2$ is indeed diagonal with
its diagonal components being proportional to 
the hierarchical fermion masses multiplied by the common factor $\zeta_f$,
see Eq.~(\ref{eq:hfalign}).
For an alternative Yukawa alignment in which ${\cal H}_2$ can couple to
light fermions sizably while still achieving the absence of
tree-level FCNCs, see Ref.~\cite{Egana-Ugrinovic:2019dqu}.}%
\begin{equation}
{\bf y}_2^f = \zeta_f\, {\bf y}_1^f\,,
\label{eq:aligned_yukawa_matrices}
\end{equation}
along with introducing the three complex alignment 
parameters $\zeta_{f=u,d,e}$.
In this case, the two aligned Yukawa matrices ${\bf y}_1^f$ 
and ${\bf y}_2^f$ 
can be diagonalized simultaneously and the Yukawa matrices
describing the couplings of $\varphi_2$ and $a$ 
fields to the fermion mass eigenstates are given by
\begin{equation}
\label{eq:hfalign}
{\bf h}_f=\zeta_f\,\frac{{\bf M}_f}{v}\,,
\end{equation}
which leads to the Hermitian and anti-Hermitian Yukawa matrices
\begin{equation}
{\bf h}_f^H=\real(\zeta_f)\,\frac{{\bf M}_f}{v}\,, \ \ \
{\bf h}_f^A=i\,\imag(\zeta_f)\,\frac{{\bf M}_f}{v}\,.
\end{equation}
When $\imag(\zeta_f)=0$, the conventional 2HDMs
based on the Glashow-Weinberg condition~\cite{Glashow:1976nt}
can be obtained
by choosing $\zeta_f$ as shown in Table~\ref{tab:2hdmtype}.
Otherwise, the couplings of the mass eigenstates of the
neutral Higgs bosons $H_{i=1,2,3}$ to two fermions are given by
\begin{equation}
-{\cal L}_{H_i\bar f f}=\sum_{i=1}^3 \sum_{f=u,d,c,s,t,b,e,\mu,\tau}
\frac{m_f}{v}\,
\bar f\left(g^S_{H_i\bar f f} + i g^P_{H_i\bar f f} \gamma_5\right) f\,H_i
\end{equation}
with the scalar and pseudoscalar couplings given by
\begin{eqnarray}
\label{eq:gsp}
g^S_{H_i\bar f f} &=& O_{\varphi_1 i} +\real(\zeta_f) O_{\varphi_2 i}
\pm \imag(\zeta_f) O_{a i}\,,\nonumber \\
g^P_{H_i\bar f f} &=& \hphantom{O_{\varphi_1 i} +}\,
\imag(\zeta_f) O_{\varphi_2 i}
\mp \real(\zeta_f) O_{a i}\,,
\end{eqnarray}
where the upper and lower signs are for the up-type fermions
$f=u,c,t$ and the down-type fermions $f=d,s,b,e,\mu,\tau$, 
respectively.
The simultaneous existence of the scalar $g^S_{H_i\bar f f}$
and pseudoscalar $g^P_{H_i\bar f f}$ couplings for a specific $H_i$
signals the CP violation in the neutral Higgs sector.
We figure out that there are two different sources of
the neutral Higgs-sector CP violation:
$(i)$ one is the CP-violating mixing among the CP-even and CP-odd states
arising in the presence of non-vanishing $\imag(Z_{5,6})$ in the Higgs potential and
$(ii)$
the other one is the complex alignment parameters of $\zeta_f$'s.
Note that the second source is absent in the conventional
four types of 2HDMs since $\zeta_f$'s are 
real in those models.

%
\begin{table}[!t]
\caption{\label{tab:2hdmtype}
Classification of the conventional
2HDMs satisfying the Glashow-Weinberg condition
\cite{Glashow:1976nt}
which guarantees the absence of
tree-level Higgs-mediated
flavor-changing neutral current (FCNC).
For the four types of 2HDM, we follow the conventions found in,
for example, Ref.~\cite{Cheung:2013rva}.
}
\setlength{\tabcolsep}{2.5ex}
\renewcommand{\arraystretch}{1.3}
\begin{center}
\begin{tabular}{|l|c|c|c|c|}
\hline
& 2HDM I   & 2HDM II
& 2HDM III & 2HDM IV \\
\hline
$\zeta_u$  & $1/t_\beta$ & $1/t_\beta$ & $1/t_\beta$ & $1/t_\beta$  \\
$\zeta_d$  & $1/t_\beta$ & $-t_\beta$ & $1/t_\beta$ & $-t_\beta$  \\
$\zeta_e$  & $1/t_\beta$ & $-t_\beta$ & $-t_\beta$ & $1/t_\beta$  \\
\hline
&
$\zeta_d=\zeta_e=\zeta_u$ &
$\zeta_d=\zeta_e=-1/\zeta_u$ &
$\zeta_d=-1/\zeta_e=\zeta_u$ &
$\zeta_d=-1/\zeta_e=-1/\zeta_u$  \\
\hline
\end{tabular}
\end{center}
\end{table}

The couplings of charged Higgs bosons to two fermions are given by
\begin{equation}
-{\cal L}_{H^\pm \bar f_\uparrow f_\downarrow}=
-\sqrt{2}\left[\overline{u_R}({\bf h}_u^\dagger V) d_L\right]H^+
+\sqrt{2}\left[\overline{u_L}(V{\bf h}_d) d_R\right]H^+
+\sqrt{2}\left[\overline{\nu_L}\,{\bf h}_e\, e_R\right]H^+
\ + \ {\rm h.c.}\,.
\end{equation}
in terms of the CKM matrix $V$ and the $3\times 3$ Yukawa matrices
${\bf h}_{u,d,e}$.

Previously, we note that the Higgs potential given by Eq.~(\ref{eq:VHiggs})
is invariant under the phase rotation 
${\cal H}_2 \to {\rm e}^{i\zeta} {\cal H}_2$ if the complex potential parameters
are accordingly rephased, see Eq.~(\ref{eq:rephaseZ567}). This observation extends
to the Yukawa interactions, Eq.~(\ref{eq:yukawa0}),
by noting that they are invariant under the phase
rotations:
\begin{eqnarray}
\label{eq:rephaseYukawa}
{\cal H}_2 \to  {\rm e}^{+ i\zeta}\,{\cal H}_2\,; \ \
{\bf y}^u_2 \to {\rm e}^{+ i\zeta} {\bf y}^u_2\,, \ \
{\bf y}^d_2 \to {\rm e}^{- i\zeta} {\bf y}^d_2\,, \ \
{\bf y}^e_2 \to {\rm e}^{- i\zeta} {\bf y}^e_2\,.
\end{eqnarray}
Under the alignment
assumption ${\bf y}_2^f = \zeta_f\, {\bf y}_1^f$ given by
Eq.~(\ref{eq:aligned_yukawa_matrices}), the above rephasing invariant rotations
become
\begin{eqnarray}
\label{eq:rephaseZeta}
{\cal H}_2 \to  {\rm e}^{+ i\zeta}\,{\cal H}_2\,; \ \
\zeta_u \to {\rm e}^{+ i\zeta} \zeta_u\,, \ \
\zeta_d \to {\rm e}^{- i\zeta} \zeta_d\,, \ \
\zeta_e \to {\rm e}^{- i\zeta} \zeta_e\,, \ \
\end{eqnarray}
in terms of the complex alignment parameters.
Then one may be able to show that the scalar and pseudoscalar couplings 
given by Eq.~(\ref{eq:gsp}) are invariant under the phase rotations 
given by Eq.~(\ref{eq:rephaseZeta}) as they should be. 
To be explicit, we first note that,
under the phase rotation ${\cal H}_2 \to {\rm e}^{i\zeta} {\cal H}_2$,
the electroweak Higgs basis changes as follow:
\begin{equation}
\left(\begin{array}{c} \varphi_1 \\ \varphi_2 \\ a \end{array}\right)
\to \ O_\zeta^T \ 
\left(\begin{array}{c} \varphi_1 \\ \varphi_2 \\ a \end{array}\right)
\ \ \ {\rm with} \ \ \
O_\zeta = \left(\begin{array}{ccc}
1 & 0 & 0 \\
0 & c_\zeta & s_\zeta \\
0 & -s_\zeta & c_\zeta
\end{array} \right)\,,
\end{equation}
which leads to
\begin{equation}
O \ \to \ O_\zeta^T \ O\,; \ \ \
{\cal M}_0^2 \to O_\zeta^T  {\cal M}_0^2 O_\zeta
\end{equation}
by observing that 
$(H_1,H_2,H_3)^T = O^T (\varphi_1,\varphi_2,a)^T$ and
${\rm diag}(M_{H_1}^2,M_{H_2}^2,M_{H_3}^2)=O^T {\cal M}_0^2 O$
should remain the same, respectively.
Under the transformation $O  \to  O_\zeta^T  O$,
the components of the mixing matrix $O$ change into
\begin{equation}
\label{eq:rephaseOmix}
O_{\varphi_1 i}  \to  O_{\varphi_1 i} \,,  \ \ \
O_{\varphi_2 i}  \to  c_\zeta O_{\varphi_2 i} - s_\zeta O_{ai} \,, \ \ \
O_{a i}  \to  s_\zeta O_{\varphi_2 i} + c_\zeta O_{ai} \,.
\end{equation}
On the other hand, under the rotations 
$\zeta_f \to {\rm e}^{\pm i\zeta} \zeta_f$
given in Eq.~(\ref{eq:rephaseZeta}), one may have
\begin{equation}
\label{eq:rephaseReInZeta}
\real({\zeta_f}) \to c_\zeta \real({\zeta_f})  \mp s_\zeta \imag({\zeta_f})\,, \ \ \
\imag({\zeta_f}) \to c_\zeta \imag({\zeta_f})  \pm s_\zeta \real({\zeta_f})\,,
\end{equation}
with the upper and lower signs being for the up-type
massive fermions $f=u$ and the down-type massive fermions
$f=d,e$, respectively. 
Using Eqs.~(\ref{eq:rephaseOmix}) and (\ref{eq:rephaseReInZeta}), it is
straightforward to show that the scalar and pseudoscalar couplings 
given by Eq.~(\ref{eq:gsp}) are invariant under the phase rotations 
of Eq.~(\ref{eq:rephaseZeta}).

To summarize, assuming ${\bf y}_2^f = \zeta_f\, {\bf y}_1^f$
with $\zeta_{f=u,d,e}$ being the three complex alignment parameters 
and combining Eqs.~(\ref{eq:rephaseZ567}) and (\ref{eq:rephaseZeta}),
we note that the Higgs potential {\it and} the Yukawa
interactions are  invariant under 
the following phase rotations:
\begin{eqnarray}
&&
{\cal H}_2 \to  {\rm e}^{+i\zeta}\,{\cal H}_2\,;\nonumber\\  &&
Y_3 \to Y_3\,{\rm e}^{-i\zeta}\,, \,
Z_5 \to Z_5\,{\rm e}^{-2i\zeta}\,, \,
Z_6 \to Z_6\,{\rm e}^{-i\zeta}\,, \,
Z_7 \to Z_7\,{\rm e}^{-i\zeta}\,;\nonumber \\ && \,
\zeta_u \to \zeta_u\,{\rm e}^{+i\zeta}\,,\,
\zeta_d \to \zeta_d\,{\rm e}^{-i\zeta}\,,\,
\zeta_e \to \zeta_e\,{\rm e}^{-i\zeta}\,,
\end{eqnarray}
which, taking account of the CP odd tadpole condition $Y_3+Z_6\,v^2/2=0$,
lead to {\it five} rephasing-invariant CPV phases in total. This leaves us more
freedom to choose the input parameters for the Higgs potential
other than
${\cal I}_{\rm CPV}$ (\ref{eq:ICPV0}),
${\cal I}^\prime_{\rm CPV}$ (\ref{eq:ICPV1}), or
${\cal I}^{\prime\prime}_{\rm CPV}$ (\ref{eq:ICPV2}).
For example, one may assign three CPV phases to the Higgs potential and
take $\zeta_u$ real and positive definite. In this case, the full set
of input parameter is to be
\begin{equation}
\label{eq:ICPVfull}
\left.{\cal I}^{V_{\cal H}\oplus{\rm Yukawa}}_{\rm CPV}
\right|_{\rm \zeta_u>0\,,\imag(\zeta_u)=0}=
\left\{v;M_{H^\pm},M_{H_1},M_{H_2},M_{H_3},\gamma,
\eta,\omega;Z_3;Z_2,Z_7\right\} \ \oplus \
\left\{|\zeta_u|,\zeta_d,\zeta_e\right\}\,,
\end{equation}
which contains 12 and 5 real degrees of freedom in the Higgs potential and
the Yukawa interactions, respectively,
with $Z_7$, $\zeta_d$ and $\zeta_e$ being fully complex.

\subsection{Interactions with Massive Vector Bosons}

The cubic interactions of the neutral and charged Higgs bosons
with the massive gauge bosons $Z$ and $W^\pm$
are described by the three interaction Lagrangians:
\begin{eqnarray}
{\cal L}_{HVV} & = & g\,M_W \, \left(W^+_\mu W^{- \mu}\ + \
\frac{1}{2c_W^2}\,Z_\mu Z^\mu\right) \, \sum_i \,g_{_{H_iVV}}\, H_i
\,,\nonumber\\[3mm]
{\cal L}_{HHZ} &=& \frac{g}{2c_W} \sum_{i>j} g_{_{H_iH_jZ}}\, Z^{\mu}
(H_i\, \!\stackrel {\leftrightarrow} {\partial}_\mu H_j) \,, \nonumber\\[3mm]
{\cal L}_{HH^\pm W^\mp} &=& -\frac{g}{2} \, \sum_i \, g_{_{H_iH^+
W^-}}\, W^{-\mu} (H_i\, i\!\stackrel{\leftrightarrow}{\partial}_\mu
H^+)\, +\, {\rm h.c.}\,,
\end{eqnarray}
respectively, where $X\stackrel{\leftrightarrow}{\partial}_\mu Y
=X\partial_\mu Y -(\partial_\mu X)Y$, $i,j =1,2,3$ and the normalized couplings
$g_{_{H_iVV}}$, $g_{_{H_iH_jZ}}$ and $g_{_{H_iH^+
W^-}}$ are given in terms of the neutral Higgs-boson $3\times 3$
mixing matrix $O$ by (note that det$(O)=\pm1$ for any orthogonal matrix $O$):
\begin{eqnarray}
\label{eq:2hdmhvvetc}
g_{_{H_iVV}} &=& O_{\varphi_1 i}
\, ,\nonumber \\
g_{_{H_iH_jZ}} &=& {\rm sign} [{\rm det}(O)] \, \, \epsilon_{ijk}\,
g_{_{H_kVV}}\, = {\rm sign} [{\rm det}(O)] \, \, \epsilon_{ijk}\, O_{\varphi_1 k}\,,
\nonumber \\
g_{_{H_iH^+ W^-}} &=& -O_{\varphi_2 i} + i O_{ai}  \, ,
\end{eqnarray}
leading to the following sum rules:
\begin{equation}
\label{eq:2hdmsumrule}
\sum_{i=1}^3\, g_{_{H_iVV}}^2\ =\ 1\,\quad{\rm and}\quad
g_{_{H_iVV}}^2+|g_{_{H_iH^+ W^-}}|^2\ =\ 1\,\quad {\rm for~ each}~
i=1,2,3\,.
\end{equation}
On the other hand, the quartic interactions of
the neutral and charged Higgs bosons
with the massive gauge bosons $Z$ and $W^\pm$ and
massless photons are given by
\begin{equation}
{\cal L}_{HHVV}=\frac{1}{v^2}\left(M_W^2 W_\mu^+ M^{\mu -} +\frac{M_Z^2}{2} Z_\mu Z^\mu
\right) \sum_{i,j=1}^3 g_{_{H_iH_jVV}} H_i H_j\,,
\end{equation}
with $g_{_{H_i H_j VV}}=\delta_{ij}$ and
\begin{eqnarray}
{\cal L}_{H^+ H^- V V} & = &
\left(
\frac{g^2}{2} W_\mu^+W^{\mu -}
+ \frac{g_Z^2 c_{2W}^2}{4} Z^\mu Z_\mu  + e^2 A^\mu A_\mu
+ {e\, g_Z\, c_{2W}^2} A^\mu Z_\mu \right)
H^+ H^-\,, \nonumber \\
{\cal L}_{H^\pm H ZW^\mp} & = & \frac{g_Z\,g\, s_W^2}{2}  \left(
Z_\mu W^{-\,\mu} \sum_{i=1}^3  g_{_{Z W^- H^+ H_i}}  H^+ H_i
+{\rm h.c.} \right)\,, \nonumber \\
{\cal L}_{H^\pm H AW^\mp} & = & -\frac{e\, g}{2} \left(
 A_\mu W^{-\,\mu} \sum_{i=1}^3 g_{_{A W^- H^+ H_i}}  H^+ H_i + {\rm h.c.}
\right)\,,
\end{eqnarray}
with $g_{_{ZW^-H^+H_i}} = g_{_{A W^- H^+ H_i}}
= - O_{\varphi_2 i}\: - i O_{a i}$,
$c_{2W}=\cos2\theta_W$, and $g_Z=g/c_W=e/(s_Wc_W)$.

\section{Constraints}
In this Section, we consider the  perturbative unitarity (UNIT) conditions and
those for the Higgs potential to be bounded from below (BFB) to obtain the
primary theoretical constraints on the potential parameters 
or, equivalently, the constraints on the Higgs-boson masses including 
correlations among them and the mixing among the three neutral Higgs bosons.
We further consider the constraints on the Higgs masses and their couplings with
vector bosons taking into account
the electroweak oblique corrections to the so-called $S$ and $T$
parameters.
We emphasize that
all the three types of constraints from the perturbative unitarity,
the Higgs potential bounded from below,
and the electroweak precision observables (EWPOs)
are independent of the basis chosen and 
working in the Higgs basis does not invoke any restrictions.
We also consider the constraints on 
$|\zeta_e|$,  $|\zeta_u|$, and the product of $\zeta_u\zeta_d$
taking account of the charged Higgs contributions to 
the flavor-changing $\tau$ decays into light leptons, 
$Z\to b\bar b$, $\epsilon_K$, and $b\to s\gamma$
\cite{Jung:2010ik,Jung:2010ab}.
\footnote{We refer to, for example,
Ref.~\cite{Crivellin:2013wna} for
an extensive study of flavor observables in the conventional
2HDMs taking the $\Phi$ basis.}

\subsection{Perturbative Unitarity}
For the unitarity conditions, we closely follow Ref.~\cite{Jurciukonis:2018skr}
\footnote{We keep our conventions for the potential parameters.}
considering the three scattering matrices of ${\cal M}_{1,2,3}^S$
which are expressed in terms of the quartic couplings $Z_{1-7}$,
see also Ref.~\cite{Kanemura:2015ska}.
The two $4\times 4$ real and symmetric
scattering matrices ${\cal M}_1^S$ and ${\cal M}_2^S$ are given by
\begin{eqnarray}
{\cal M}_1^S &=& \left( \begin{array}{cc}
\eta_{00}-I & \eta^T \\
\eta & E+I\times {\bf 1}_{3\times 3} \end{array} \right)\,; \ \ \
{\cal M}_2^S = \left( \begin{array}{cc}
3\eta_{00}-I & 3\eta^T \\
3\eta & 3E+I\times {\bf 1}_{3\times 3} \end{array} \right)\,,
\end{eqnarray}
where $\eta_{00}=Z_1+Z_2+Z_3$ and $I=Z_3-Z_4$.
The row vector $\eta^T$ is given by
\begin{eqnarray}
\eta^T&=&(\real(Z_6+Z_7),-\imag(Z_6+Z_7),Z_1-Z_2)\,,
\end{eqnarray}
and the $3\times 3$ real and symmetric matrix $E$ by
\begin{eqnarray}
E &=& \left( \begin{array}{ccc}
Z_4 +2\real(Z_5) & -2\imag(Z_5) & \real(Z_6-Z_7) \\
-2\imag(Z_5) & Z_4-2\real(Z_5) & -\imag(Z_6-Z_7) \\
\real(Z_6-Z_7) & -\imag(Z_6-Z_7) & Z_1+Z_2-Z_3 \end{array} \right)\,.
\end{eqnarray}
The third $3\times 3$ scattering matrix ${\cal M}_3^S$ 
is Hermitian which takes the form of
\begin{eqnarray}
{\cal M}_3^S &=& \left( \begin{array}{ccc}
2Z_1 & 2Z_5 & \sqrt{2}Z_6 \\
2Z_5^* & 2Z_2 & \sqrt{2}Z_7^* \\
\sqrt{2}Z_6^* & \sqrt{2}Z_7 & Z_3+Z_4 \end{array} \right)\,.
\end{eqnarray}
And then, the unitarity conditions are imposed by requiring
that the 11 eigenvalues of the
three scattering matrices ${\cal M}_{1,2,3}^S$ and the quantity $I$ should have
their moduli smaller than $4\pi$.
%

When $Z_6=Z_7=0$, the 12 unitarity conditions simplify into
\begin{eqnarray}
\label{eq:unt1}
\left| Z_3 \pm Z_4 \right| &<& 4\pi\,, \nonumber \\[2mm]
\left| Z_3 \pm 2|Z_5| \right| &<& 4\pi\,, \nonumber \\[2mm]
\left| Z_3 +2Z_4 \pm 6|Z_5| \right| &<& 4\pi\,, \nonumber \\[2mm]
\left| Z_1 +Z_2 \pm \sqrt{(Z_1-Z_2)^2+4|Z_5|^2} \right| &<& 4\pi\,, \nonumber \\[2mm]
\left| Z_1 +Z_2 \pm \sqrt{(Z_1-Z_2)^2+Z_4^2} \right| &<& 4\pi\,, \nonumber \\[2mm]
\left| 3Z_1 +3Z_2 \pm \sqrt{9(Z_1-Z_2)^2+(2Z_3+Z_4)^2} \right| &<& 4\pi\,.
\end{eqnarray}
While taking $Z_1=Z_2=Z_3=Z_4=Z_5=0$, one may have
\begin{eqnarray}
\label{eq:unt2}
\sqrt{|Z_6|^2+|Z_7|^2} &<& 2\sqrt{2}\pi\,,  \ \ \
\sqrt{|Z_6|^2+|Z_7|^2+|Z_6^2+Z_7^2|} < \frac{4\pi}{3}\,.
\end{eqnarray}
Then, by combining them, one may arrive at the following
UNIT conditions for individual parameters~\cite{Jurciukonis:2018skr}
\begin{eqnarray}
\label{eq:untindiv}
&&
|Z_{1,2,5}|<2\pi/3\,, \ \ \ |Z_{6,7}| <2\sqrt{2}\pi/3\,, \nonumber \\[2mm]
&&
|Z_3-Z_4|<4\pi \ \cup \
|2Z_3+Z_4|<4\pi \ \cup \
|Z_3+2Z_4|<4\pi \,.
\end{eqnarray}

\subsection{Higgs Potential Bounded-from-below}
We consider the following 5 necessary conditions
for the most general 2HDM Higgs potential 
with explicit CP violation
to be bounded-from-below in a marginal sense
\cite{Jurciukonis:2018skr}:
\footnote{
Denoting the quartic part of the scalar potential as $V_4$,
a marginal stability requirement means that
$V_4\geq 0$ for any
direction in field space tending to infinity~\cite{Branco:2011iw}.
In contrast, a strong stability requirement is $V_4>0$ without
the equality sign. In this work,
we adopt the marginal stability requirement.
}
\begin{eqnarray}
\label{eq:bfb}
Z_1 \geq 0 \,, \ \ \ Z_2 \geq 0\,;&& \nonumber \\[2mm]
2\sqrt{Z_1Z_2}+Z_3 \geq  0 \,, \ \ \  2\sqrt{Z_1Z_2}+Z_3+Z_4-2|Z_5|\geq 0\,;&& \nonumber \\[2mm]
Z_1+Z_2+Z_3+Z_4+2|Z_5|-2|Z_6+Z_7|  \geq 0\,.&&
\end{eqnarray}
Note that though the quartic couplings $Z_2$ and $Z_7$ have no direct relations
to the masses and mixing of Higgs bosons but
they are interrelated with the other five quartic couplings of $Z_{1,3-6}$
through the UNIT and BFB conditions.

\subsection{Electroweak Precision Observables}

The electroweak oblique corrections to the so-called $S$, $T$ and $U$
parameters~\cite{Peskin:1990zt,Peskin:1991sw} provide significant
constraints on the quartic couplings of the 2HDM.
Fixing $U=0$ which is suppressed by
an additional factor $M_Z^2/M^2_{\rm BSM}$
\footnote{Here, $M_{\rm BSM}$ denotes some heavy mass scale involved
with new physics beyond the Standard Model.} compared to
$S$ and $T$,
the $S$ and $T$ parameters are constrained as follows
\begin{equation}
\label{eq:STRange}
\frac{(S-\widehat S_0)^2}{\sigma_S^2}\ +\
\frac{(T-\widehat T_0)^2}{\sigma_T^2}\ -\
2\rho_{ST}\frac{(S-\widehat S_0)(T-\widehat T_0)}{\sigma_S \sigma_T}\
\leq\ R^2\,(1-\rho_{ST}^2)\; ,
\end{equation}
with $R^2=2.3$, $4,61$, $5.99$, $9.21$, $11.83$ at $68.3 \%$, $90 \%$,
$95 \%$, $99 \%$, and $99.7 \%$  confidence levels (CLs), respectively.
For our numerical analysis, we adopt the 95\% CL limits.
The central values and their standard deviations are given by
\footnote{See the 2020 edition of the review
``{\it\bf 10. Electroweak Model and Constraints on New Physics}"
by J.Erler and A. Freitas in
Ref.~\cite{ParticleDataGroup:2020ssz}.}
\begin{equation}
\label{eq:STPDG}
(\widehat S_0\,,\ \sigma_S)\ =\ (0.00\,,\ 0.07)\;,\qquad
(\widehat T_0\,,\ \sigma_T)\ =\ (0.05 \,,\ 0.06)\; ,
\end{equation}
with a strong correlation $\rho_{ST}=0.92$ between
$S$ and $T$ parameters.
The electroweak oblique parameters, which are defined to arise
from new physics only, are in excellent agreement with
the SM values of zero for the reference values
of $M_{H_{\rm SM}}=125.25$ GeV and $M_t=172.5$ GeV
\cite{ParticleDataGroup:2020ssz}.

In 2HDM, the $S$ and $T$ parameters might be estimated using
the following expressions~\cite{Toussaint:1978zm,Lee:2012jn}
\begin{eqnarray}
  \label{eq:STphi}
S_\Phi \!&=&\! -\frac{1}{4\pi} \left[
\left(1+\delta_{\gamma Z}^{H^\pm}\right)^2F^\prime_\Delta(M_{H^\pm},M_{H^\pm})
-\sum_{(i,j)=(1,2)}^{(1,3),(2,3)}
\left(g_{_{H_iH_jZ}}+\delta_Z^{H_iH_j}\right)^2 F^\prime_\Delta(M_{H_i},M_{H_j})
\right]\,, \\
T_\Phi \!&=&\! -\frac{\sqrt{2}G_F}{16\pi^2\alpha_{\rm EM}}\ \left[
-\sum_{i=1}^{3}\left|g_{_{H_iH^- W^+}}+\delta_W^{H_i}\right|^2F_\Delta(M_i,M_{H^\pm})
+\sum_{(i,j)=(1,2)}^{(1,3),(2,3)}
\left(g_{_{H_iH_jZ}}+\delta_Z^{H_iH_j}\right)^2 F_\Delta(M_{H_i},M_{H_j})
\right]\; .\quad   \nonumber
\end{eqnarray}
In this work, we ignore the vertex corrections
$\delta_{\gamma Z}^{H^\pm}$, $\delta_Z^{H_iH_j}$, and
$\delta_W^{H_i}$ since the size of
the most of the quartic couplings are smaller than $3$  and
the quantum corrections
proportional to $\sim Z_i^2/16\pi^2$ might be negligible. 
Then, we observe that
all the relevant couplings are determined by the
three physical couplings of $g_{_{H_iVV}}$ since
$g_{_{H_iH_jZ}}^2 = |\epsilon_{ijk}| g_{_{H_kVV}}^2
=|\epsilon_{ijk}| O_{\varphi_1 k}^2$ and
$|g_{_{H_iH^- W^+}}|^2=1-g_{_{H_iVV}}^2=1-O_{\varphi_1 i}^2$.
The one-loop functions are given by
\footnote{See, for example, Ref.\cite{Kanemura:2011sj}.}
\begin{eqnarray}
\label{eq:ffp}
F_\Delta(m_0,m_1) &=& F_\Delta(m_1,m_0) =
\frac{m_0^2+m_1^2}2 -\frac{m_0^2m_1^2}{m_0^2-m_1^2}\ln\frac{m_0^2}{m_1^2}\,,
\nonumber \\[3mm]
F_\Delta^\prime(m_0,m_1) &=& F_\Delta^\prime(m_1,m_0) =
-\frac{1}{3} \left[ \frac{4}{3}
-\frac{m_0^2 \ln m_0^2 -m_1^2 \ln m_1^2}{m_0^2-m_1^2}
-\frac{m_0^2+m_1^2}{(m_0^2-m_1^2)^2}F_\Delta(m_0,m_1) \right]\,.
\end{eqnarray}
We note that $F_\Delta(m,m)=0$ and $F_\Delta^\prime(m,m)=\frac{1}{3}\ln m^2$.
\footnote{Here and after,
$\ln m^2$ could be understood as, for example,
$\ln\left[m^2/(1\,{\rm GeV})^2\right]$ if necessary. }
When $g_{_{H_1VV}}^2=1$,
neglecting the $Z^2$-dependent vertex correction factors
$\delta_{\gamma Z}^{H^\pm}$, $\delta_W^{H_i}$  and $\delta_Z^{H_iH_j}$,
$S_\Phi$ and $T_\Phi$ are symmetric under the exchange $M_{H_2}\leftrightarrow M_{H_3}$
and they are identically vanishing when $M_{H_2}=M_{H_3}=M_{H^\pm}$.
\footnote{The $S_\Phi$ and $T_\Phi$ parameters are
independent of $M_{H_1}$ when $g_{_{H_1VV}}^2=1$.}

\subsection{Flavor Constraints on the Alignment Parameters}

The alignment parameters $\zeta_{f=u,d,e}$ are constrained by considering
the charged Higgs contributions to the low energy observables such as
flavor-changing $\tau$ decays,
leptonic and semileptonic decays of pseudoscalar mesons,
the $Z\to b \bar b$ process,
$B$ meson mixing,
the CPV parameter $\epsilon_K$ in $K$ meson mixing, and
the radiative $b\to s\gamma$ decay~\cite{Jung:2010ik}.
In this work, we consider the 
flavor constraints on  the absolute sizes
of $\zeta_{e}$, $\zeta_{u}$ and $\zeta_{d}$.
Note that we neglect the constraints on the
products of the alignment parameters
taking account of only the single constraints
on the absolute values of
$\zeta_e$, $\zeta_u$ and $\zeta_d$ under the assumption
that they are fully independent from each other.

The flavor-changing $\tau$ decays into light leptons provide the 
following constraint on $|\zeta_{e}|$~\cite{Jung:2010ik}:
\begin{equation}
\label{eq:upperZetae}
|\zeta_e|\leq 200 \ \left(\frac{M_{H^\pm}}{500\,{\rm GeV}}\right)\,,
\end{equation}
at 95\% CL.
On the other hand, the constraint on $|\zeta_{u}|$ may come from 
the $Z$-peak precision observables involving the $Z\to b\bar b$ decay assuming
the quantum corrections to the $Zb\bar b$ vertex beyond the SM is dominated by
the charged Higgs contributions.
More explicitly, the ratio $R_b=\Gamma(Z\to b\bar b)/\Gamma(Z\to\,{\rm hadrons})$
is used by neglecting the contributions depending on $|\zeta_{d}|$ which
are suppressed by the factor $\overline{m}_t(M_Z)/\overline{m}_b(M_Z) \sim 60$
compared to those depending on $|\zeta_{u}|$.
It turns out that
the upper limit on $|\zeta_{u}|$ linearly increases with $M_{H^\pm}$ as follow
\cite{Jung:2010ik}:
\begin{equation}
\label{eq:upperZetau}
|\zeta_u|\leq 0.72 + 1.19 \left(\frac{M_{H^\pm}}{500\,{\rm GeV}}\right)
\ \ \ (95\%\ {\rm CL})\,.
\end{equation}
To be very strict, the above upper limit should be applied 
only when $|\zeta_{d}|=0$.
The similar while more direct upper limit $|\zeta_{u}|$
could be obtained by considering
the CPV parameter $\epsilon_K$ in $K$ meson mixing which depends
on $|\zeta_{u}|$ only neglecting the masses of the light $d$ and $s$ quarks.
Actually the limit from $\epsilon_K$ is 
slightly stronger than that from $Z\to b \bar b$ by the amount of 
about 10\%~\cite{Jung:2010ik}.
In this work, for the upper limit on $|\zeta_{u}|$,
we apply the slightly weaker constraint from $Z\to b \bar b$ given by
Eq.~(\ref{eq:upperZetau})
while considering it valid independently of $\zeta_{d}$.
In passing, for the $\Delta B=2$ processes mediated by
box diagrams with exchanges of $W^\pm$ and/or $H^\pm$ bosons,
we note that the leading Willson coefficients 
which are not suppressed by the light quark mass
depend $\zeta_u$ and $\zeta_d$. 
When $\zeta_d=0$, one might obtain the similar
upper limit on $|\zeta_u|$ as that from $\epsilon_K$
\cite{Jung:2010ik}.

\begin{figure}[t!]
\vspace{-1.0cm}
\begin{center}
\includegraphics[width=10.5cm]{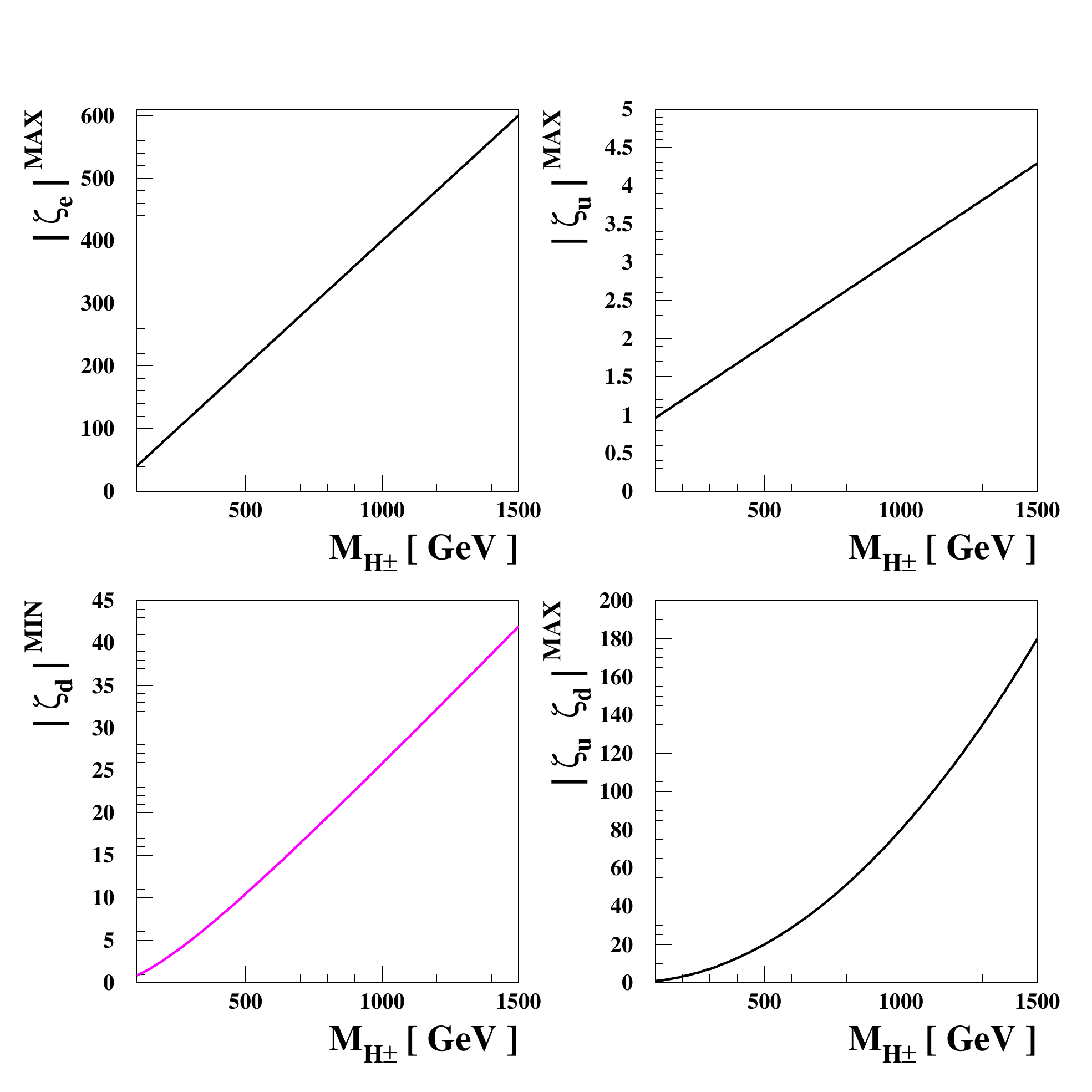}
\end{center}
\vspace{-0.5cm}
\caption{\it 
The 95\% CL limits on the alignment parameters 
as functions of $M_{H^\pm}$
obtained by considering
the charged Higgs contributions to low energy 
observables~\cite{Jung:2010ik}.
(Upper Left) The upper limit on $|\zeta_e|$ from flavor-changing
$\tau$ decays into light leptons, see Eq.~(\ref{eq:upperZetae}).
(Upper Right) The upper limit on $|\zeta_u|$ from
$R_b$ and $\epsilon_K$, see Eq.~(\ref{eq:upperZetau}).
(Lower Left)  The minimum value of $|\zeta_d|$ required to satisfy the
$b\to s\gamma$ constraint
through the destructive interference when
$\zeta_u\zeta_d$ is positive, see Eq.~(\ref{eq:lowerZetad}).
The $R_b$ and $\epsilon_K$ constraints on $|\zeta_u|$ are combined.
(Lower Right) The upper limit on the product 
of $|\zeta_u|$ and $|\zeta_d|$ from
$b\to s\gamma$ when  $\zeta_u$ and $\zeta_d$  are complex,
see Eq.~(\ref{eq:upperZetauZetad}).
}
\label{fig:x1234}
\end{figure}

There is no limit on $\zeta_d$ independently of $\zeta_u$ and/or $\zeta_e$. 
But one may extract some interesting information on $\zeta_d$ 
considering the radiative $b\to s\gamma$ decay. 
Numerically, the decay amplitude can be cast into the following form
\cite{Jung:2010ab,Borzumati:1998tg}:
\footnote{
Note that the product $\zeta_u\zeta_d$ is the rephasing 
invariant quantity in our convention, see Eq.~(\ref{eq:rephaseZeta}).}
\begin{equation}
{\cal A} \sim {\cal A}_{\rm SM}\left\{
1-0.1\, \zeta_u\zeta_d \left(\frac{500\ {\rm GeV}}{M_{H^\pm}}\right)^2
+ 0.01\, |\zeta_u|^2 \left(\frac{500\ {\rm GeV}}{M_{H^\pm}}\right)^2
\right\}\,.
\end{equation}
When $\zeta_u\zeta_d$ is negative, the interference with the SM amplitude
is always constructive and the product is constrained to be
small and, as usual, $|\zeta_d|$ can be significantly larger (smaller) than 1
only when $|\zeta_u|$ is very small (large).
On the contrary,
if $\zeta_u\zeta_d$ is positive,
$|\zeta_d|$ could be large independently of $|\zeta_u|$.
In this case, a destructive
interference occurs and the experimental constraints can be satisfied
when
\begin{equation}
\label{eq:btosgamma}
\zeta_u\zeta_d \sim 20 \left(\frac{M_{H^\pm}}{500\ {\rm GeV}}\right)^2\,.
\end{equation}
Combining the upper limit on $|\zeta_u|$ given by Eq.~(\ref{eq:upperZetau}),
we observe that the destructive interference can always occur when
\begin{equation}
\label{eq:lowerZetad}
|\zeta_d| \ \gsim \ 20\ \frac{\left(\frac{M_{H^\pm}}{500\ {\rm GeV}}\right)^2}
{0.72+1.19\left(\frac{M_{H^\pm}}{500\ {\rm GeV}}\right)}\,,
\end{equation}
and $\zeta_u\zeta_d>0$.
Most generally, allowing $\zeta_u\zeta_d$ to be complex, 
it turns out that the rough 95\% CL upper limit on
the absolute value of the product 
is basically saturated by the relation
given by Eq.~(\ref{eq:btosgamma})~\cite{Jung:2010ik} or
\begin{equation}
\label{eq:upperZetauZetad}
|\zeta_u||\zeta_d| \lsim 20\
\left(\frac{M_{H^\pm}}{500\ {\rm GeV}}\right)^2
\ \ \ (95\%\ {\rm CL})\,.
\end{equation}
For the summary, we present the upper limits on 
$|\zeta_e|$, $|\zeta_u|$, and $|\zeta_e\zeta_d|$
and the lower limit on $|\zeta_d|$ in Fig.~\ref{fig:x1234}.

Before closing this section, we briefly comment on the constraints 
from the heavy Higgs boson searches carried out at the LHC.
The heavy neutral Higgs bosons have been searched through their
decays into
$\tau^+\tau^-$~\cite{ATLAS:2017eiz,CMS:2018rmh,Bailey:2020ecs,ATLAS:2020zms},
$b\bar b$~\cite{CMS:2018hir},
$t\bar t$~\cite{ATLAS:2017snw,CMS:2019rvj,ATLAS:2022ohr},
$WW$~\cite{ATLAS:2017jag},
$ZZ$~\cite{ATLAS:2017tlw,ATLAS:2017otj,CMS:2018amk,ATLAS:2020tlo},
$Zh_{\rm 125 GeV}$~\cite{ATLAS:2017xel,CMS:2019qcx}, etc.
On the other hand, the charged Higgs boson search channels include
the decay modes into
$\tau^\pm\nu$~\cite{ATLAS:2018gfm,CMS:2019bfg},
$tb$~\cite{ATLAS:2018ntn,CMS:2020imj,ATLAS:2021upq},
$cb$~\cite{CMS:2018dzl}, 
$cs$~\cite{CMS:2015yvc,CMS:2020osd},
and $Wh_{\rm 125 GeV}$~\cite{ATLAS:2017xel}.
Basically, the experimental upper limits on
the product of the production cross section and 
the decay rate into a specific search mode have been analyzed
to obtain the allowed parameter space of
a specific model.
For example, 
the search in the $\tau^+\tau^-$ final state 
excludes the presence of a heavy neutral Higgs
with $M_A$ below about 1 TeV at 95\% CL
in the minimal supersymmetric extension of the SM (MSSM)
when, depending on scenarios,
$\tan\beta \gsim 15\sim 25$ and
the exclusion contour reaches up to
$M_A=1.6$ TeV for $\tan\beta=60$~\cite{CMS:2018rmh}. 
While in the aligned 2HDM taken in this work, the Yukawa couplings
of the up- and down-type quarks and the charged leptons to heavy 
Higgs bosons are completely uncorrelated and the interpretation
of the experimental limits is much more involved.
This is because
the three alignment parameters of $\zeta_{u,d,e}$ are independent 
from each other while all of them are involved in the calculation of 
the decay rate pertinent to a specific search mode.
In principle, one can easily avoid the constraints from, for example,
$H/A \to \tau\tau$ and $H^\pm \to \tau\nu$ 
by taking $|\zeta_{e}|\ll 1$.
But it might be still allowed to have
$|\zeta_{e}| \gsim 20$ and $M_A<1$ TeV
if one can suppress the branching fraction into $\tau^+\tau^-$ 
by choosing the other alignment parameters of
$\zeta_u$ and $\zeta_d$ appropriately.
In this respect, a through analysis of the experimental search
results in the framework of aligned 2HDM with three
independent alignment parameters 
deserves an independent full consideration.
%
In this work, we simply assume that the parameter space
considered in the next Section could be made 
more or less safe 
from the LHC constraints from no observation of 
significant excess in the heavy Higgs boson searches
by judiciously manipulating the three alignment 
parameters which are otherwise uncorrelated.

\section{Numerical Analysis}
From the relation $g_{_{H_1VV}}=O_{\varphi_1 1}$
given in Eq.~(\ref{eq:2hdmhvvetc}) and the expressions for
the $H_i$ couplings
to the two SM fermions given in 
Eq.~(\ref{eq:gsp}),
one might define the Yukawa delay factor $\Delta_{H_1\bar f f}$
by the amount of which the decoupling of the Yukawa couplings
of the lightest Higgs boson is delayed
compared to its  coupling to a pair of massive vector bosons:
\begin{equation}
\label{eq:delayfactor}
\Delta_{H_1\bar f f}\equiv \sqrt{
\left(g^S_{H_1\bar f f}-g_{_{H_1VV}}\right)^2+
\left(g^P_{H_1\bar f f}\right)^2} = |\zeta_f|\left(1-g_{_{H_1VV}}^2\right)^{1/2}\,,
\end{equation}
where we use the relation $\sum_{\alpha=\varphi_1,\varphi_2,a}O_{\alpha i}^2 =1$
for $i=1$.
We observe that the delay factor $\Delta_{H_1\bar f f}$ defined above
is basis-independent and can be generally used even in the CPV case.
Anticipating that the impacts on the Yukawa delay factor
due to the CP-violating phases of $Z_{5,6,7}$
and $\zeta_{u,d,e}$ are redundant,
we consider the CP-conserving (CPC) case for our numerical study for simplicity.
For a recent global analysis of the aligned CPC 2HDM
taking account of several phenomenological constraints as well as theoretical
requirements, we refer to Ref.~\cite{Eberhardt:2020dat} 
but with a caution.
\footnote{
In Ref.~\cite{Eberhardt:2020dat}, the authors 
take $\lambda_{5,6,7}$
for the fitting parameters in addition to the Higgs masses 
$m_h=125.10$ GeV, $M_H$, $M_A$, $M_{H^\pm}$, and the mixing angle
$\tilde\alpha$. In our notations, they use the set of input parameters of
$\{v;M_{H^\pm}, M_h, M_H, M_A, \gamma;Z_5,Z_6,Z_7\}$. Comparing to
${\cal I}_{\rm CPC}^\prime$ given in Eq.~(\ref{eq:input_cpc}), we find that
the potential parameters $Z_5$ and $Z_6$ are used more than needed 
while $Z_2$ and $Z_3$ are missing in the set.
Note that $Z_5$ and $Z_6$ are entirely fixed when 
the mixing angle and the three neutral Higgs masses are
given, see Eq.~(\ref{eq:Z1456_cpc}), and the parameter $Z_2$ should be included
at least because it is independent of the Higgs masses 
and mixing like as $Z_7$.}

\subsection{UNIT and BFB constraints}

\begin{figure}[t!]
\vspace{-1.0cm}
\begin{center}
\includegraphics[width=8.0cm]{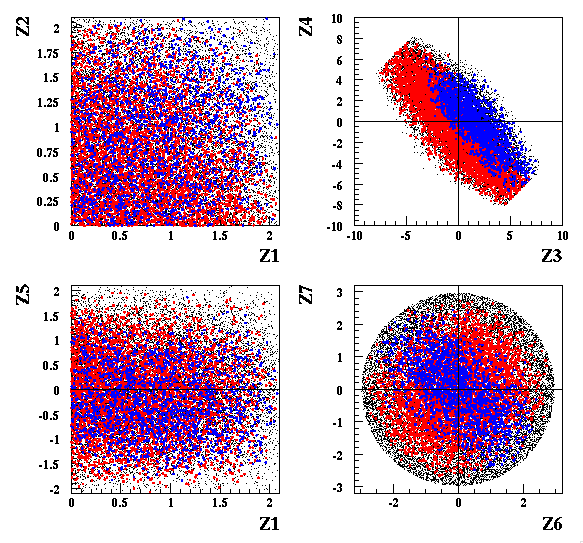}
\includegraphics[width=8.0cm]{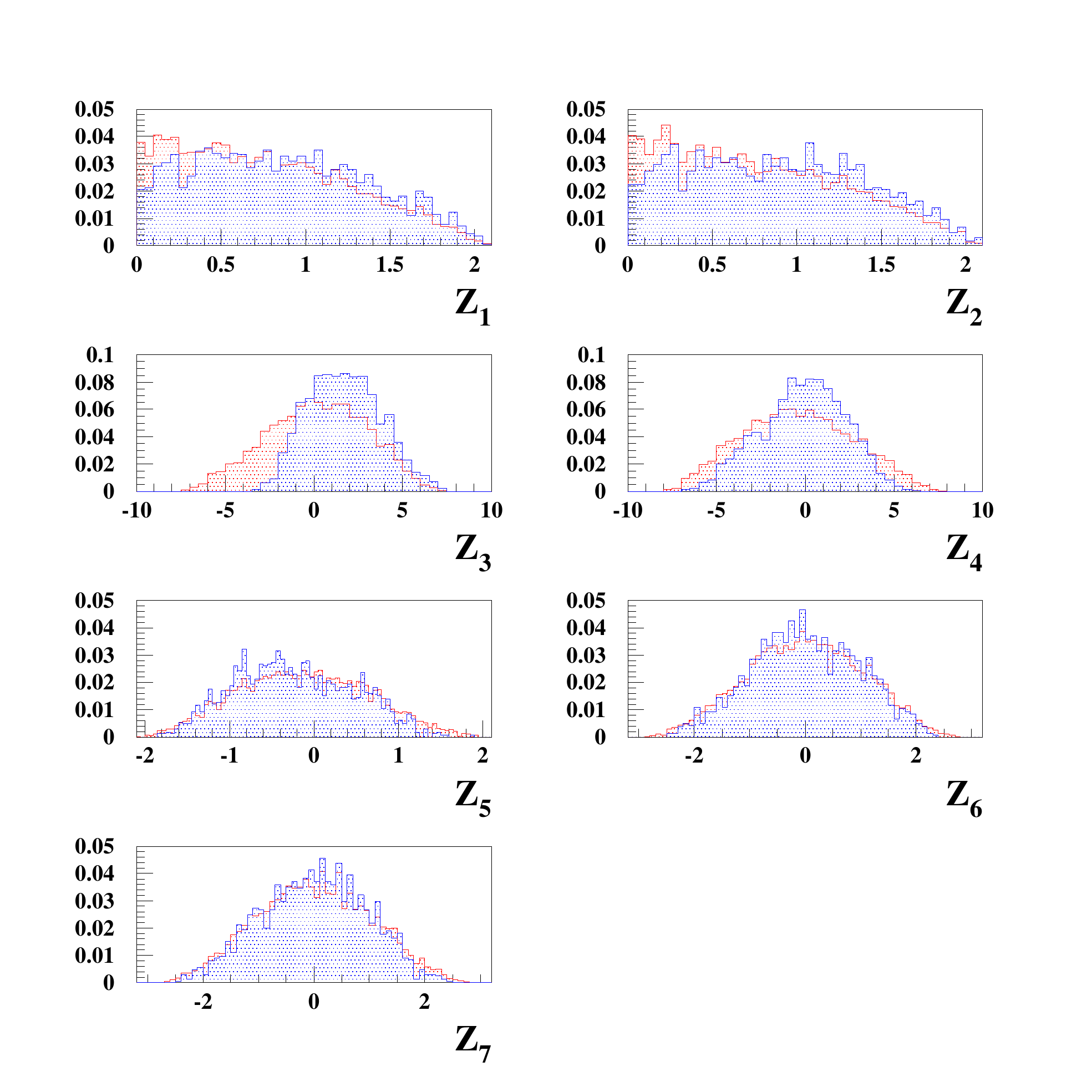}
\end{center}
\vspace{-0.5cm}
\caption{\it The {\bf UNIT} and {\bf BFB} constraints using ${\cal I}^Z_{\rm CPC}$,
see Eq.~(\ref{eq:izcpc}):
(Left) Scatter plots (red) of 
$Z_2$ versus $Z_1$ (upper left),
$Z_4$ versus $Z_3$ (upper right), 
$Z_5$ versus $Z_1$ (lower left), and
$Z_7$ versus $Z_6$ (lower right)
with the UNIT conditions imposed.
For the blue points, the necessary BFB conditions are
additionally imposed.
Also shown are the points in black
which are obtained by requiring only the simplified UNIT conditions
in Eqs.~(\ref{eq:unt1}) and (\ref{eq:unt2}).
(Right) The normalized
distributions of the quartic couplings obtained by requiring
only the UNIT (red) and the combined UNIT$\oplus$BFB conditions.}
\label{fig:unitbfb}
\end{figure}

First of all, we consider the UNIT and BFB constraints.
Observing that the two conditions depend only on the quartic couplings
$Z_{1-7}$, we take the following set of input parameters:
\footnote{To have ${\cal I}^Z_{\rm CPC}$ from Eq.~(\ref{eq:y2zi}), we trade
$M_{H^\pm}$ with $Z_3$. Note that the dimensionful parameter $Y_2$
is irrelevant for the UNIT and BFB constraints.}
\begin{eqnarray}
\label{eq:izcpc}
{\cal I}^Z_{\rm CPC} &=&
\left\{v,Y_2;Z_1,Z_2,Z_3,Z_4,Z_5,Z_6,Z_7\right\}\,.
\end{eqnarray}
In the left panel of Fig.~\ref{fig:unitbfb},
we show the scatter plots of
$Z_2$ versus $Z_1$ (upper left),
$Z_4$ versus $Z_3$ (upper right),
$Z_5$ versus $Z_1$ (lower left), and
$Z_7$ versus $Z_6$ (lower right).
The plots are produced by randomly generating the
quartic couplings in the ${\cal I}^Z_{\rm CPC}$ set.
In each plot, the black points are obtained
by imposing only the simplified UNIT conditions of
Eqs.~(\ref{eq:unt1}) and (\ref{eq:unt2}). The full consideration
of the UNIT conditions based on the scattering
matrices ${\cal M}^S_{1,2,3}$ produces the red points.
The results
obtained by simultaneously imposing the full UNIT and BFB conditions
(UNIT$\oplus$BFB) are denoted by the blue points.
We find our results are very consistent with
those presented in Ref.~\cite{Jurciukonis:2018skr}.
After imposing the UNIT and UNIT$\oplus$BFB conditions,
we note that the normalized distributions of the quartic
couplings are no longer flat as shown in the right panel
of Fig.~\ref{fig:unitbfb}. As in the left panel,
the distributions of the quartic couplings obtained by requiring
only the UNIT (red) conditions and the combined UNIT$\oplus$BFB conditions
are in red and blue, respectively.
We note that the smaller $|Z_6+Z_7|$ and the
positive $Z_3$ values are preferred by further imposing the BFB conditions
in addition to the UNIT ones, see Eq.~(\ref{eq:bfb}).

\subsection{Electroweak constraints}

Coming to the electroweak (ELW) constraints, since the oblique
corrections are expressed in terms of the masses and couplings of
Higgs bosons, it is more natural and
convenient to take the following set of input parameters:
\begin{eqnarray}
\label{eq:iprimcpc}
{\cal I}_{\rm CPC}^\prime &=&
\left\{v;M_{H^\pm},M_h=M_{H_1},M_H,M_A,\gamma ;Z_3;Z_2,Z_7\right\}\,,
\end{eqnarray}
referring to Eq.~(\ref{eq:input_cpc}).
In the ${\cal I}_{\rm CPC}^\prime$ set, all the
massive parameters are physical Higgs masses except
$v = \left(\sqrt{2}G_F\right)^{-1/2} \simeq 246.22$ GeV.
We assume that the neutral state $h=H_1$ is the lightest Higgs boson
and plays the role of the SM Higgs boson in the decoupling limit of $s_\gamma=0$
by taking $M_{H_1}=125.5$ GeV~\cite{Sirunyan:2020xwk}.
And, for the masses of heavy Higgs bosons,
we randomly generate their masses squared between
$M_{H_1}^2$ and $(1.5~{\rm TeV})^2$.
For the mixing angle $\gamma$, we take the convention of $|\gamma|\leq \pi/2$
without loss of generality resulting in
$c_\gamma \geq 0$ and ${\rm sign}(s_\gamma)={\rm sign}(Z_6)$.
For the implementation of the UNIT and BFB constraints using the set
${\cal I}_{\rm CPC}^\prime$, we recall 
the quartic couplings $Z_{1,4,5,6}$ 
in terms of the Higgs masses and the 
mixing angle $\gamma$ in the CPC case  
given by Eq.~(\ref{eq:Z1456_cpc}).

Using the set ${\cal I}^\prime_{\rm CPC}$ for the input parameters
in the CPC case, the $S$ and $T$ parameters given by
Eq.~(\ref{eq:STphi}) 
take the following simpler forms:
\begin{eqnarray}
\label{eq:STCPC}
S_{\Phi}^{\rm CPC} \!&=&\! -\frac{1}{4\pi} \left[
F^\prime_\Delta(M_{H^\pm},M_{H^\pm})
-c_\gamma^2\,F^\prime_\Delta(M_A,M_H)
-s_\gamma^2\,F^\prime_\Delta(M_A,M_h)
\right]\,, \nonumber\\
T_{\Phi}^{\rm CPC} \!&=&\! \frac{\sqrt{2}G_F}{16\pi^2\alpha_{\rm EM}}\ [
F_\Delta(M_A,M_{H^\pm})+c_\gamma^2\,F_\Delta(M_H,M_{H^\pm})+
s_\gamma^2\,F_\Delta(M_h,M_{H^\pm}) \nonumber \\
&& \hspace{4.6cm}
-c_\gamma^2\,F_\Delta(M_A,M_H) -s_\gamma^2\,F_\Delta(M_A,M_h)
]\; ,\quad
\end{eqnarray}
ignoring the vertex corrections. We observe that
$T_{\Phi}^{\rm CPC}$ is identically vanishing when $M_{H^\pm}=M_A$
and, when $M_{H^\pm}\sim M_A \sim M_H \gg M_h$, we obtain
\footnote{For $S_{\Phi}$, note that
$[\ln M_A^2/3+(M_H-M_A)/3M_A]-[\ln M_H^2/3+(M_A-M_H)/3M_H]
\simeq (M_H-M_A)^3/9M_A^3$.}
\begin{eqnarray}
\label{eq:STapp}
S_{\Phi}^{\rm CPC} &\simeq& -\frac{1}{4\pi}\left[
\frac{\ln M_{H^\pm}^2}{3}
-c_\gamma^2\, \left(\frac{\ln M_A^2}{3}+\frac{M_H-M_A}{3M_A}\right)
-s_\gamma^2\, \left(\frac{\ln M_A^2}{3}-\frac{5}{18}\right)\right] \,, \nonumber \\[2mm]
T_{\Phi}^{\rm CPC} &\simeq& \frac{\sqrt{2}G_F}{16\pi^2\alpha_{\rm EM}}\left[
\frac{2(M_A-M_{H^\pm})^2}{3}
+c_\gamma^2\,\frac{2(M_H-M_{H^\pm})^2}{3}
+s_\gamma^2\, \frac{M_{H^\pm}^2}{2} \right. \nonumber \\
&& \hspace{4.9cm} \left.
-\,c_\gamma^2\,\frac{2(M_A-M_H)^2}{3} \
-s_\gamma^2\, \frac{M_A^2}{2}\right]\,,
\end{eqnarray}
keeping the leading terms.
To obtain Eq.~(\ref{eq:STapp}) for
the approximated expressions of the $S$ and $T$ parameters, we use 
\begin{eqnarray}
F_\Delta(m_0,m_1) &=& \frac{2(m_0-m_1)^2}{3} -\frac{(m_0-m_1)^4}{30\,m_1^2}
+{\cal O}\left[\frac{(m_0-m_1)^5}{m_1^3}\right]\,, \nonumber \\[2mm]
F^\prime_\Delta(m_0,m_1) &=& \frac{\ln m_1^2}{3}
+\frac{(m_0-m_1)}{3m_1} -\frac{(m_0-m_1)^2}{30\, m_1^2}
+{\cal O}\left[\frac{(m_0-m_1)^3}{m_1^3}\right]\,,
\end{eqnarray}
for $m_0\sim m_1$ and 
\begin{eqnarray}
F_\Delta(m_0,m_1) &=& \frac{m_1^2}{2}
+\left(\frac{1}{2}+\ln\frac{m_0^2}{m_1^2}\right)\,m_0^2
+{\cal O}\left[\left(\frac{m_0^4}{m_1^2}\right)
\ln\frac{m_0^2}{m_1^2}\right]\,, \nonumber \\[2mm]
F^\prime_\Delta(m_0,m_1) &=& \frac{\ln m_1^2}{3}
-\frac{5}{18} +\frac{2}{3}\frac{m_0^2}{m_1^2}
+{\cal O}\left[\left(\frac{m_0^4}{m_1^4}\right)
\ln\frac{m_0^2}{m_1^2}\right]\,,
\end{eqnarray}
for $m_1\gg m_0$.

\begin{figure}[t!]
\vspace{-1.0cm}
\begin{center}
\includegraphics[width=8.5cm]{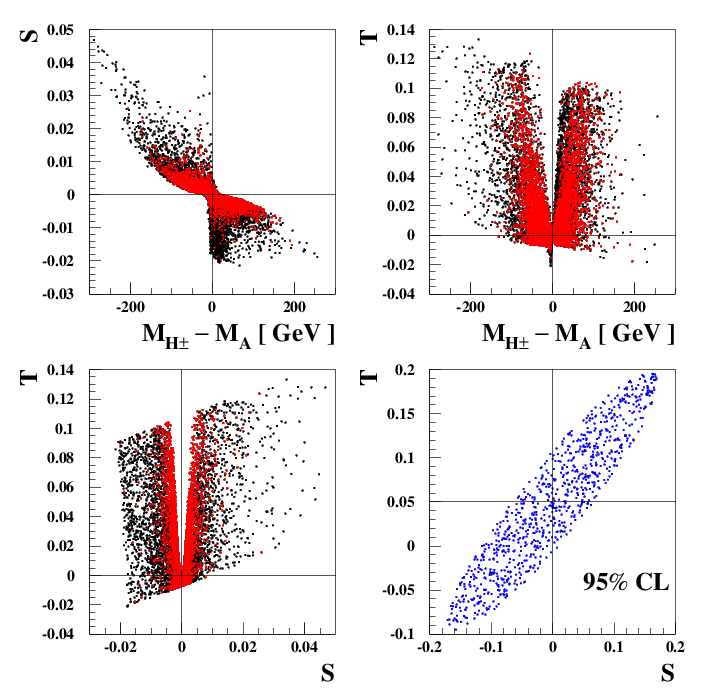}
\includegraphics[width=8.5cm]{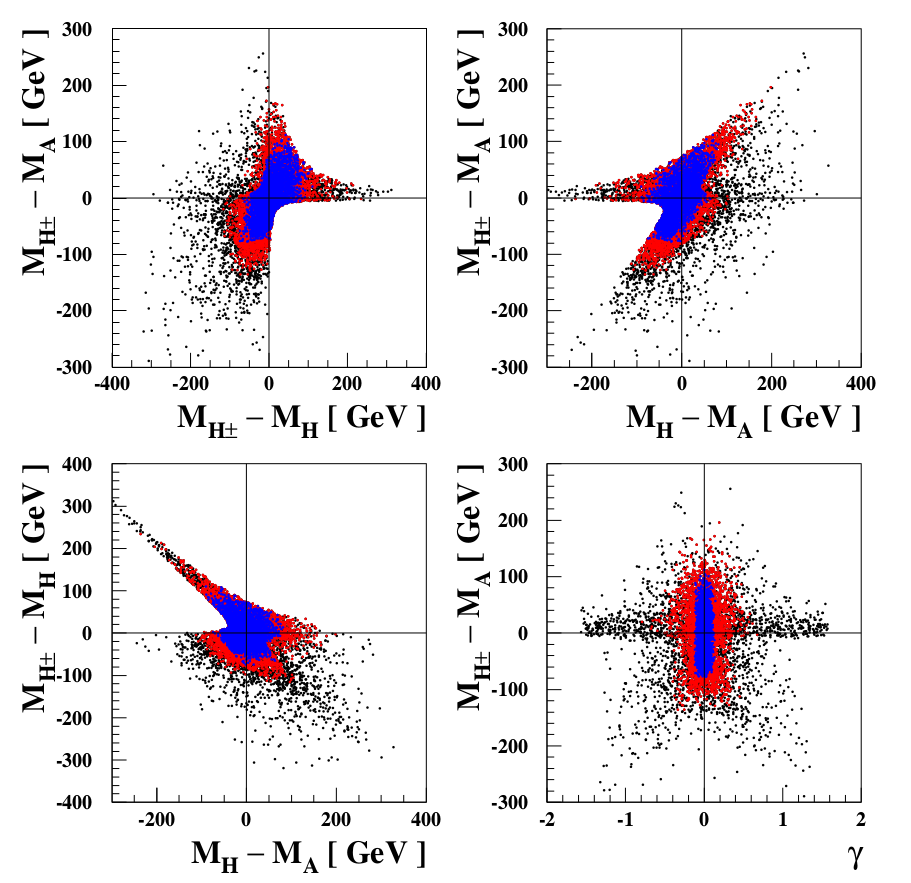}
\end{center}
\vspace{-0.5cm}
\caption{\it
Correlations among the $S$ and $T$ parameters, mass differences, and the mixing
angle $\gamma$ using the set ${\cal I}^\prime_{\rm CPC}$.
(Left) 
Scatter plots of
$S$ versus $M_{H^\pm}-M_A$ (upper left),
$T$ versus $M_{H^\pm}-M_A$ (upper right), and
$T$ versus $S$ (lower left)
with the combined UNIT$\oplus$BFB$\oplus$ELW$_{95\%}$ constraints
imposed (black). The red points are with the
small angle condition $|\gamma|<0.1$.
In the lower-right plot, as a reference,
the 95\% CL ELW constraint on the $S$ and $T$
parameters according to Eqs.~(\ref{eq:STRange}) and (\ref{eq:STPDG}) 
is shown.
(Right) Scatter plots of
$M_{H^\pm}-M_A$ versus $M_{H^\pm}-M_H$ (upper left),
$M_{H^\pm}-M_A$ versus $M_{H}-M_A$ (upper right),
$M_{H^\pm}-M_H$ versus $M_{H}-M_A$ (lower left), and
$M_{H^\pm}-M_A$ versus $\gamma$ (lower right)
with the combined UNIT$\oplus$BFB$\oplus$ELW$_{95\%}$ constraints
imposed (black). The red and blue points are for
$M_{H^\pm}>500$ GeV and $M_{H^\pm}>1$ TeV, respectively.}
\label{fig:stmm}
\end{figure}
In the left panel of Fig.~\ref{fig:stmm}, we show the $S$ and $T$
parameters imposing the UNIT, BFB, and ELW constraints abbreviated by
the combined UNIT$\oplus$BFB$\oplus$ELW$_{95\%}$ ones.
Note that the 95\% CL ELW limits are adopted and
the heavy Higgs masses squared are scanned up to $(1.5~{\rm TeV})^2$.
We find that $S$ takes values in the range
between $-0.02$ and $0.05$ whose
absolute values are smaller than $\sigma_S=0.07$, see Eq.~(\ref{eq:STPDG}).
Actually, we find that $|S|<\sigma_S$ even with only
the UNIT and BFB constraints imposed.
Note that $S$ is mostly negative (positive) when $M_{H^\pm} > (<) M_A$.
Specifically, we find that $S \simeq -1/4\pi\,(5/18) \simeq -0.02$
when $M_{H^\pm}-M_A =0$ and $\gamma=\pi/2$.
The $T$ parameter takes its value
between $-0.02$ and $0.13$ which are given by the delimited
range determined by $-0.02<S<0.05$,
the strong correlation $\rho_{ST}=0.92$ and $R_{95\%}^2=5.99$,
see Eqs.~(\ref{eq:STRange}) and (\ref{eq:STPDG}) and the lower-right plot
in the left panel of Fig.~\ref{fig:stmm}.
Incidentally, we observe that $T=0$ when $M_{H^\pm}=M_A$ though it quickly
deviates from $0$  when $M_{H^\pm} \neq M_A$.
In the right panel of Fig.~\ref{fig:stmm}, we show the
correlations among the mass differences and the mixing
angle $\gamma$ using the set ${\cal I}^\prime_{\rm CPC}$.
We find that
\begin{eqnarray}
&&
|M_H-M_A|/{\rm GeV}\lsim 200\,(100) \,, \ \ \
|M_{H^\pm}-M_H|/{\rm GeV}\lsim 200\,(110) \,, \nonumber \\
&&
|M_{H^\pm}-M_A|/{\rm GeV}\lsim 200\,(110) \,, \ \ \
|\gamma|\lsim  0.8\,(0.14)\,,
\end{eqnarray}
when $M_{H^\pm} \gsim$ 500 GeV (1 TeV).

\begin{figure}[t!]
\vspace{-1.0cm}
\begin{center}
\includegraphics[width=8.5cm]{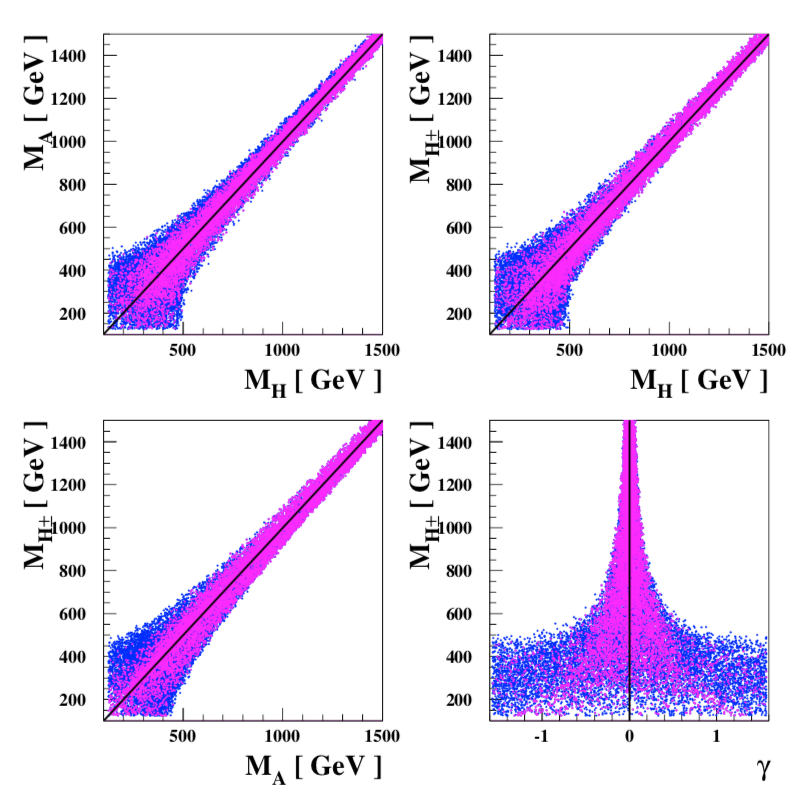}
\includegraphics[width=8.5cm]{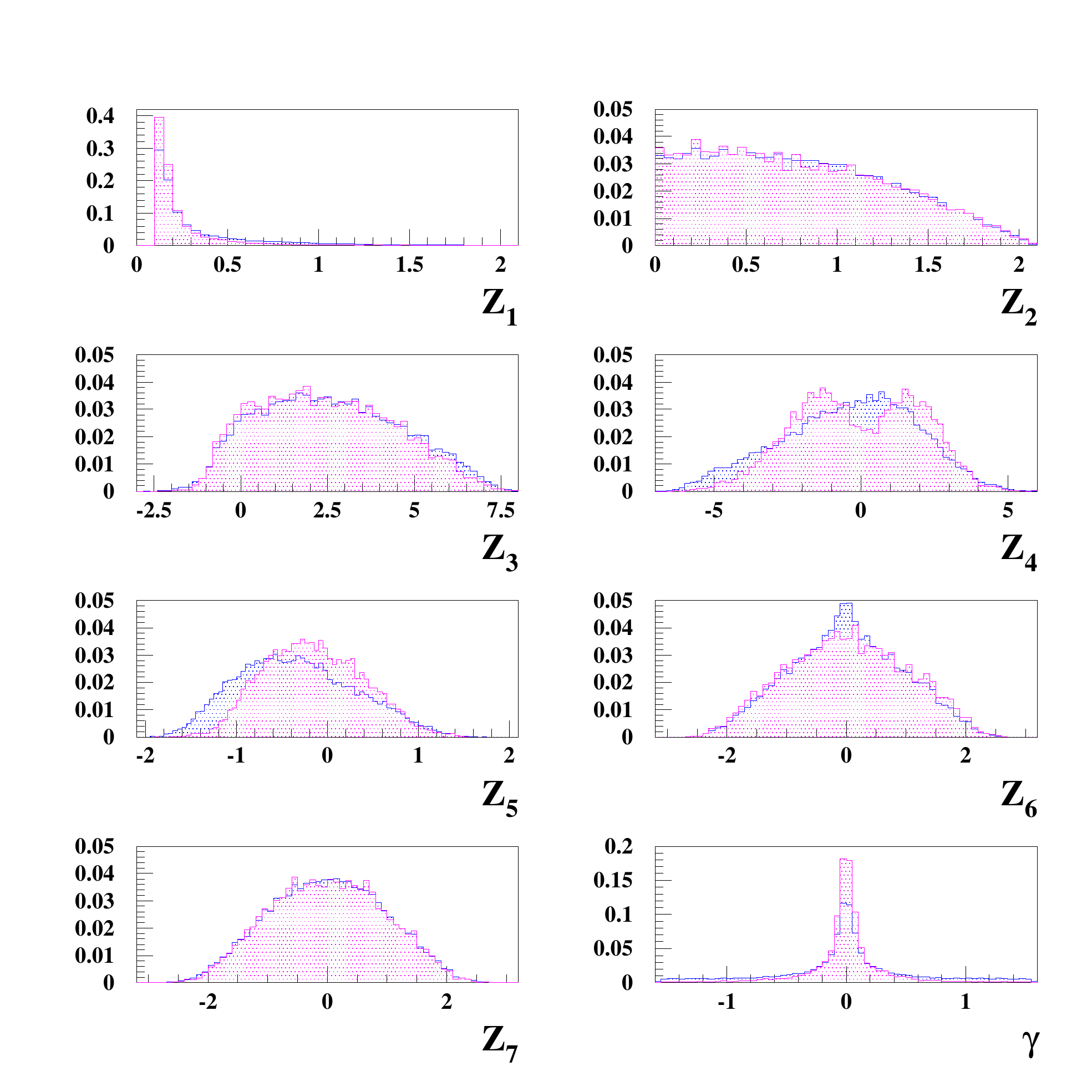}
\end{center}
\vspace{-0.5cm}
\caption{\it The UNIT$\oplus$BFB$\oplus$ELW$_{95\%}$ constraints (magenta)
using ${\cal I}^\prime_{\rm CPC}$, see Eq.~(\ref{eq:iprimcpc}).
For comparisons, we also show the results after applying
only the UNIT$\oplus$BFB constraints (blue).:
(Left) Scatter plots of $M_A$ versus $M_H$ (upper left),
$M_{H^\pm}$ versus $M_H$ (upper right),
$M_{H^\pm}$ versus $M_A$ (lower left), and
$M_{H^\pm}$ versus $\gamma$ (lower right).
%
(Right) The normalized distributions of the quartic couplings
and the mixing angle $\gamma$.
}
\label{fig:mmziang}
\end{figure}
We show the correlations among the heavy Higgs-boson masses and the mixing
angle $\gamma$ in the left panel of Fig.~\ref{fig:mmziang}.
Requiring the ELW constraint in addition to the UNIT$\oplus$BFB ones,
we find that $Z_1$ and $\gamma$ take values near to 0 
and $Z_4$ and $Z_5$ positive ones more likely,
see the right panel of Fig.~\ref{fig:mmziang}.
We find that
the UNIT and BFB conditions combined with the ELW constraint
restrict the quartic couplings as follows:
\begin{eqnarray}
&& \hspace{0.33cm}
0.1\lsim Z_1 \lsim 2.0\,, \ \ \
0  \lsim Z_2 \lsim 2.1\,, \ \ \
-2.4 \lsim Z_3 \lsim 8.0\,, \ \ \
-6.3 \lsim Z_4 \lsim 6.0\,, \nonumber \\
&&
-1.9\lsim Z_5\lsim 1.6\,, \ \ \
-2.7\lsim Z_6\lsim 2.7\,, \ \ \
-2.7\lsim Z_7\lsim 2.7\,.
\end{eqnarray}

\subsection{Alignment of Yukawa couplings}

Now, we have come to the point to address 
the alignment of Yukawa couplings. 
When we talk about the alignment of the Yukawa
couplings in general 2HDMs, we imply:
$(i)$ the alignment of them in the flavor space and
$(ii)$ the alignment of
the lightest Higgs-boson couplings
to a pair of the SM fermions in the decoupling limit
of $M_{H,A,H^\pm}\to \infty$.
By $(i)$, we precisely mean the assumption that
the two Yukawa matrices of ${\bf y}_1^{f}$ and ${\bf y}_2^{f}$
are aligned in the flavor space or
${\bf y}_2^{f} = \zeta_{f} {\bf y}_1^{f}$, see 
Eq.(\ref{eq:aligned_yukawa_matrices}),
which, in the CPC case, leads to
\begin{eqnarray}
\label{eq:gsCPC}
g^S_{H_1\bar f f} = O_{\varphi_1 1} +\zeta_f O_{\varphi_2 1}
= c_\gamma - \zeta_f s_\gamma \,,
\end{eqnarray}
with $f=u$ and $d$ for the up- and down-type quarks, respectively,
and $f=e$ for the three charged leptons.
Then, by $(ii)$, one might mean
\begin{equation}
g^S_{H_1\bar f f} \ \to \ 1 \ \ \ {\rm as} \ \ \
M_{H,A,H^\pm}\to \infty\,.
\end{equation}
In Eq.~(\ref{eq:gsCPC}), we note that
the quantity $c_\gamma$ is nothing but the coupling
$g_{_{H_1VV}}=O_{\varphi_1 1}=c_\gamma$ which is driven to take the
SM value of $1$ by the combined UNIT, BFB, and ELW constraints
as $M_{H,A,H^\pm}$ increases. 
Therefore, from Eq.~(\ref{eq:delayfactor}),
the Yukawa delay factor simplifies to
$\Delta_{H_1\bar f f}=|\zeta_f s_\gamma|$, and
the alignment of
the lightest Higgs-boson couplings to the SM fermions
in the decoupling limit
is delayed by the amount of $|\zeta_f s_\gamma|$
which can not be ignored even when $|s_\gamma| \ll 1$
if $|\zeta_f|$ is significantly larger than $1$.

\begin{figure}[t!]
\vspace{-1.0cm}
\begin{center}
\includegraphics[width=8.5cm]{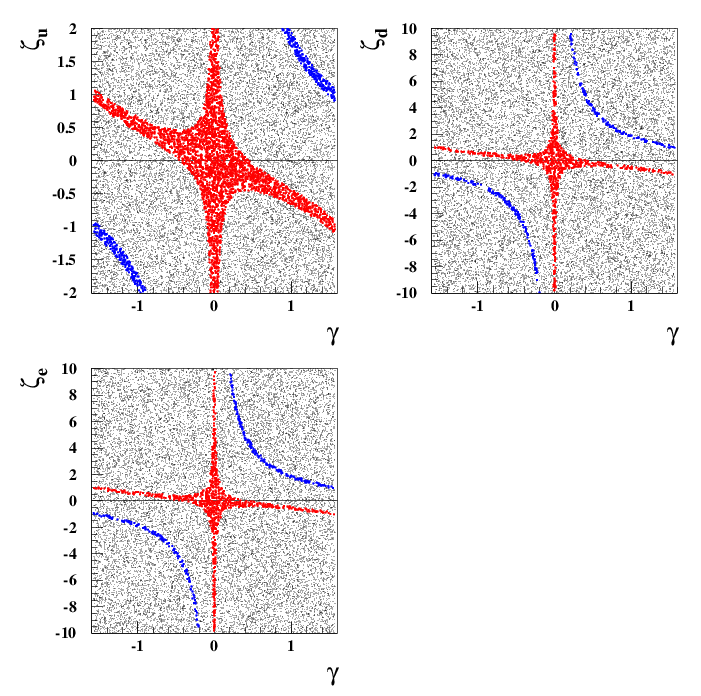}
\includegraphics[width=8.5cm]{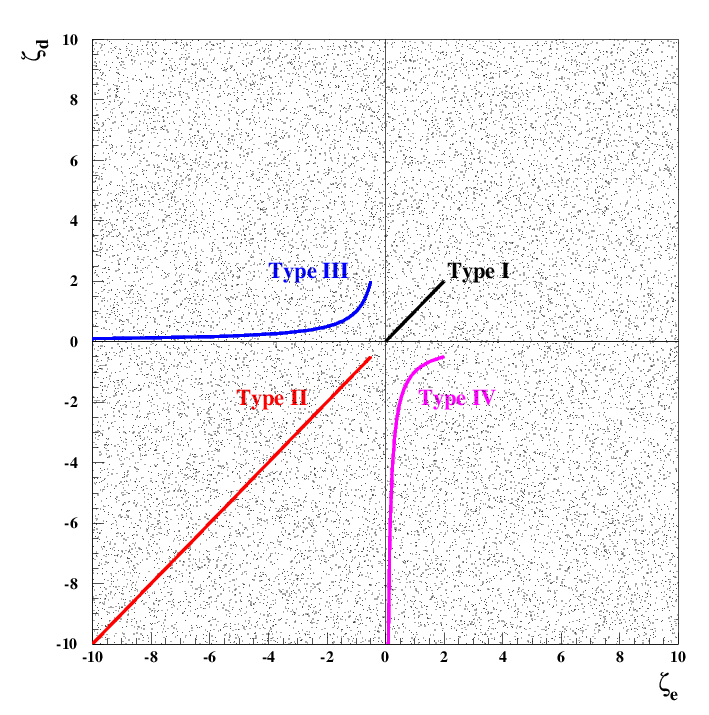}
\end{center}
\vspace{-0.5cm}
\caption{\it
(Left) Scatter plots of
$\zeta_u$ versus $\gamma$ (upper left),
$\zeta_d$ versus $\gamma$ (upper right), and
$\zeta_e$ versus $\gamma$ (lower left)
obtained by scanning $-\pi/2\leq\gamma\leq\pi/2$ and
the three real parameters of the set ${\cal I}^\zeta_{\rm CPC}$
in the ranges of $-2<\zeta_u<2$ and $-10<\zeta_{d,e}<10$.
On each $\zeta_f$-$\gamma$ plane, the regions
satisfying $|g^S_{H_1\bar f f}-1|<0.1$
and $|g^S_{H_1\bar f f}+1|<0.1$ are denoted in red and blue, respectively.
(Right) Scatter plot of $\zeta_d$ versus $\zeta_e$ with $1/100<\zeta_u<2$.
The four lines represent the four conventional 2HDMs as denoted
taking $1/2<t_\beta<100$.
}
\label{fig:zf2hdm}
\end{figure}
For a quantitative study,
in addition to ${\cal I}^\prime_{\rm CPC}$ given by Eq.~(\ref{eq:iprimcpc}),
we have added the following set of input parameters containing
three real parameters:
\begin{equation}
\label{input_zeta}
{\cal I}^{\zeta}_{\rm CPC} =\left\{\zeta_u,\zeta_d,\zeta_e\right\}\,.
\end{equation}
In the left panel of Fig.~\ref{fig:zf2hdm}, we show the
correlations between each of
the three alignment parameters $\zeta_{f=u,d,e}$
and the mixing angle $\gamma$ when the absolute value
of the corresponding coupling $g^S_{H_1\bar f f}$ is within
10\% range of the SM value of $1$ or
$|g^S_{H_1\bar f f}-1|<0.1$ and
$|g^S_{H_1\bar f f}+1|<0.1$ for
$g^S_{H_1\bar f f}>0$ (red) and $g^S_{H_1\bar f f}<0$ (blue), respectively.
Scanning $|\gamma|\leq\pi/2$, $g^S_{H_1\bar f f} \simeq 1$ near $\gamma=0$.
At $\gamma=\pm \pi/2$, the $g^S_{H_1\bar f f}$ coupling
takes the value of $1$ when $\zeta_f=\mp 1$ (red).
While if $\zeta_f=\pm 1$, we note that $g^S_{H_1\bar f f}=-1$ 
at $\gamma=\pm \pi/2$ (blue).
In the right panel of Fig.~\ref{fig:zf2hdm}, by the four lines, we show the
correlations between $\zeta_d$ and $\zeta_e$ in the
four conventional 2HDMs
\footnote{The parameters $\zeta_d$ and $\zeta_e$ are completely
uncorrelated in the general 2HDM based on the relation
Eq.~(\ref{eq:aligned_yukawa_matrices}) as shown by the scattered
black dots in the right panel of Fig.~\ref{fig:zf2hdm}.}
based on appropriately defined discrete $Z_2$ symmetries
taking $1/100<\zeta_u=1/t_\beta<2$, see Table~\ref{tab:2hdmtype}.
We observe that both $\zeta_d$ and $\zeta_e$
are bounded {\it only} in the type-I 2HDM
between $1/100$ and $2$.
Otherwise, at least one of them is limitless in principle.
Therefore, except the type-I 2HDM,
$g^S_{H_1\bar d d}$ and/or $g^S_{H_1\bar e e}$ could be
largely deviated from 1 in the decoupling limit
even when $\zeta_u$ is limited.

\begin{figure}[t!]
\begin{center}
\includegraphics[width=8.5cm]{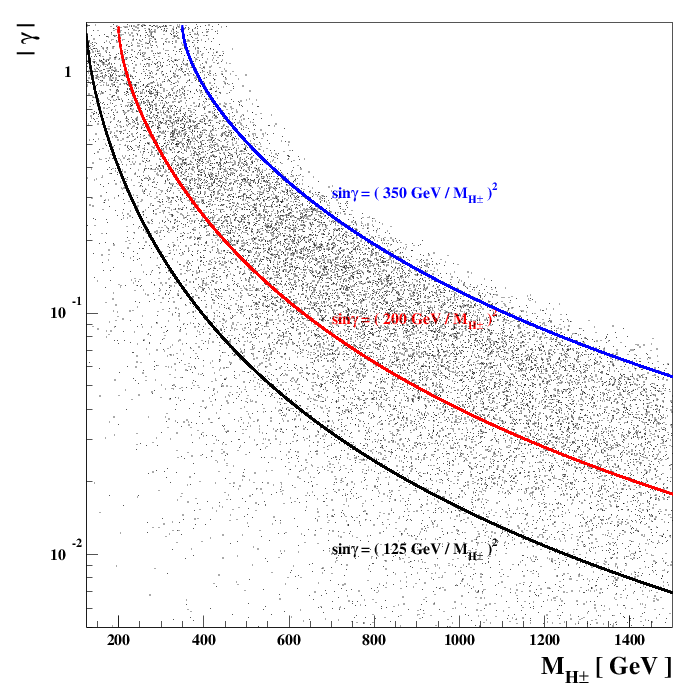}
\includegraphics[width=8.5cm]{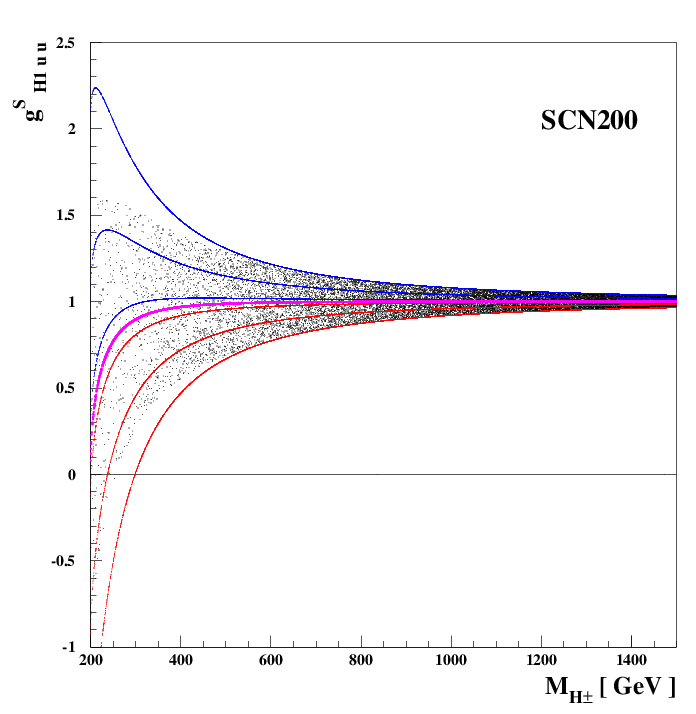}
\includegraphics[width=8.5cm]{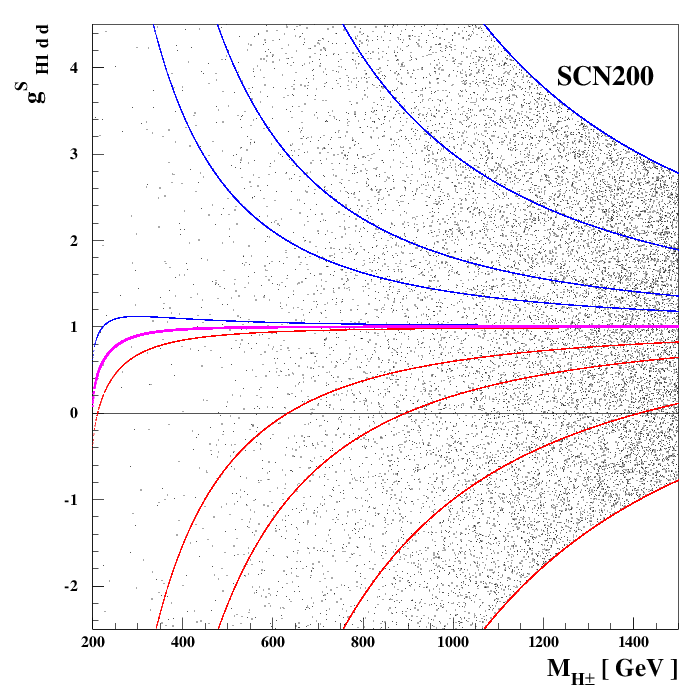}
\includegraphics[width=8.5cm]{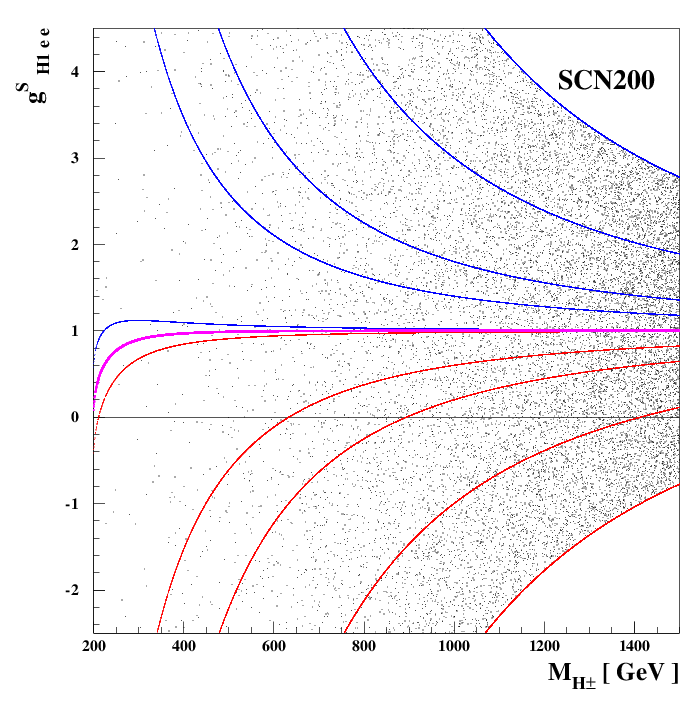}
\end{center}
\vspace{-0.5cm}
\caption{\it
(Upper Left)
The same as in the lower-right plot in the left panel of Fig.~\ref{fig:mmziang}
but for $|\gamma|$ versus $M_{H^\pm}$
with the UNIT$\oplus$BFB$\oplus$ELW$_{95\%}$ constraints imposed.
The three curves show the cases of
$|\sin\gamma| = (125~{\rm GeV}/M_{H^\pm})^2$ (black),
$|\sin\gamma| = (200~{\rm GeV}/M_{H^\pm})^2$ (red), and
$|\sin\gamma| = (350~{\rm GeV}/M_{H^\pm})^2$ (blue) from bottom to top.
(Upper Right) Scatter plot of
$g^S_{H_1\bar u u}$ versus $M_{H^\pm}$ taking {\bf SCN200}
in which
the upper limit on $|\zeta_u|$ from $R_b$ and $\epsilon_K$
is applied, see Eq.~(\ref{eq:upperZetau}).
The blue (red) lines are for
$\cos\gamma +(-) \zeta_u |\sin\gamma|$ for $\zeta_u=2,1,0.2$ from the
outermost lines to the magenta one which is for $g_{_{H_1VV}}=\cos\gamma$.
(Lower Left) Scatter plot of
$g^S_{H_1\bar d d}$ versus $M_{H^\pm}$ taking {\bf SCN200}.
The blue (red) lines are for
$\cos\gamma +(-) \zeta_d|\sin\gamma|$ for $\zeta_d=100,50,20,10,0.5$ from the
outermost lines to the magenta one which is for $g_{_{H_1VV}}=\cos\gamma$.
(Lower Right)
The same as in the lower-left plot but for $g^S_{H_1\bar e e}$ versus $M_{H^\pm}$
with the lines for
$\cos\gamma \pm \zeta_e|\sin\gamma|$ for $\zeta_e=100,50,20,10,0.5$.
}
\label{fig:gammch_t5}
\end{figure}

To concentrate on the alignment of
the lightest Higgs-boson couplings
to a pair of the SM fermions in the decoupling limit
of $M_{H,A,H^\pm}\to \infty$ under the assumption of
${\bf y}_2^{f} \propto {\bf y}_1^{f}$ as in
Eq.~(\ref{eq:aligned_yukawa_matrices}),
we consider a simplified scenario in which the mixing
angle $|\sin\gamma|$ is inversely proportional to $1/M_{H^\pm}^2$
reflecting the behavior of $|\sin\gamma|=|g_{_{HVV}}|$ being
suppressed by the quartic powers of the heavy
Higgs-boson masses at leading order~\cite{Choi:2020uum}.
In the upper-left frame of Fig.~\ref{fig:gammch_t5}, we show
the scatter plot for $|\gamma|$ versus $M_{H^\pm}$ together with
the three curves showing the cases of
$\sin\gamma = (125~{\rm GeV}/M_{H^\pm})^2$ (black),
$\sin\gamma = (200~{\rm GeV}/M_{H^\pm})^2$ (red), and
$\sin\gamma = (350~{\rm GeV}/M_{H^\pm})^2$ (blue) from bottom to top.
\footnote{The choice of $\sin\gamma = (m_6/M_{H^\pm})^2$
is equivalent to fix
$Z_6=(M_H^2-m_h^2)\cos\gamma \sin\gamma/v^2 \sim (m_6/v)^2$
when $M_H \sim M_{H^\pm} \gg M_h$ and $\sin(\gamma) \ll 1$,
see Eq.~(\ref{eq:Z1456_cpc}).}
The input parameters are the same as in Fig.~\ref{fig:mmziang} and
the combined UNIT$\oplus$BFB$\oplus$ELW$_{95\%}$ constraints are imposed.
For illustration, we take the case of $\sin\gamma = (200~{\rm GeV}/M_{H^\pm})^2$.
The coupling of the lightest Higgs boson $H_1$ to a pair of massive vector bosons
are constrained by the precision LHC Higgs data~\cite{Cheung:2018ave}.
We note that, for example, $\cos\gamma=g_{_{H_1VV}}\gsim 0.95$ or
$|\sin\gamma|=|g_{_{HVV}}|\lsim 0.3$ can be satisfied
when $M_{H^\pm}\gsim 400$ GeV for this choice.
We further assume that
the masses of the heavy Higgs bosons of $H$, $A$, and $H^\pm$ are degenerate.
This assumption reflects the fact that the combined
UNIT $\oplus$ BFB $\oplus$ ELW$_{95\%}$ constraints prefers quite degenerate
heavy-Higgs bosons when they weigh more than about 400 GeV as shown
in the left panel of Fig.~\ref{fig:mmziang}.
We dub this scenario {\bf SCN200} in which we precisely fix and vary
the input parameters in the two sets of
${\cal I}^\prime_{\rm CPC}$ and ${\cal I}^\zeta_{\rm CPC}$
as follows:
\begin{eqnarray}
\label{eq:scn200}
{\bf SCN200} \ : \ &&
{\hphantom\oplus} \{M_h=M_{H_1}=125.5 ~{\rm GeV},
M_H=M_A=M_{H^\pm}=[200..1500] ~{\rm GeV}; \nonumber \\
&&{\hphantom\oplus~~}
\sin\gamma=\pm (200~{\rm GeV}/M_{H^\pm})^2;
Z_2=[0..2], Z_3=[-3..8], Z_7=[-3..3]\} \nonumber \\
&& \oplus \{\zeta_u=[1/100..2], \zeta_d=[-100..100], \zeta_e=[-100..100]\} \,,
\end{eqnarray}
together with the combined
UNIT $\oplus$ BFB $\oplus$ ELW$_{95\%}$ constraints imposed.
In this scenario, the Yukawa delay factor is given by
\begin{equation}
\left.\Delta_{H_1\bar f f}\right|_{\bf SCN200}
=|\zeta_f|\,\frac{(200~{\rm GeV})^2}{M_{H^\pm}^2}
\simeq |\zeta_f|\,\frac{|Z_6|v^2}{M_{H^\pm}^2}\,,
\end{equation}
with $\left.Z_6\right|_{\bf SCN200}\simeq 0.66$.
Note that we use the approximation
$Z_6v^2 = (M_H^2-M_h^2)\, c_\gamma s_\gamma \simeq
M_{H^\pm}^2\, s_\gamma$ in the above equation.

In the upper-right frame of Fig.~\ref{fig:gammch_t5}, we show the scatter plot
of $g^S_{H_1\bar u u}$ versus $M_{H^\pm}$ taking {\bf SCN200}
in which
the upper limit on $|\zeta_u|$ from $R_b$ and $\epsilon_K$
is applied, see Eq.~(\ref{eq:upperZetau}).
We observe that the coupling $g^S_{H_1\bar u u}$ is within
about 30\% and 10\% ranges of the SM value of $1$ when $M_{H^\pm}>500$ GeV
and $M_{H^\pm}>1$ TeV, respectively. As previously discussed,
the alignment of the coupling $g^S_{H_1\bar f f}$ is delayed by
the amount of $\zeta_f \sin\gamma$ compared to the coupling $g_{_{H_1VV}}$ and
$g^S_{H_1\bar u u}$ is most deviated from its SM value of $1$ 
by the amount of
\begin{equation}
\left.\Delta_{H_1\bar u u}\right|_{{\rm\bf SCN200}\,,|\zeta_u|\leq 2}
=|\zeta_u|\,\left(\frac{200~{\rm GeV}}{M_{H^\pm}}\right)^2
\leq 0.32\, \left(\frac{500~{\rm GeV}}{M_{H^\pm}}\right)^2\,.
\end{equation} 
To make this point clear, we add the blue and red lines showing
$g^S_{H_1\bar u u}$ taking $\zeta_u=0.2,1$, and $2$ and
the magenta one showing $g_{_{H_1VV}}$. We indeed see that
$g^S_{H_1\bar u u}$ is most close to $g_{_{H_1VV}}$ when $\zeta_u=0.2$
and the two lines taking $\zeta_u=2$ provide the envelope which
includes all the scattered points.
\footnote{Note that the line segments for $\zeta_u=2$ 
with $M_{H^\pm}\lsim 500$ GeV are located outside the scattered region
implying that they are excluded by
the upper limit on $|\zeta_u|$ from $R_b$ and $\epsilon_K$.}
In the lower frames of Fig.~\ref{fig:gammch_t5}, the scatter plots
of $g^S_{H_1\bar d d}$ versus $M_{H^\pm}$  (left) and
$g^S_{H_1\bar e e}$ versus $M_{H^\pm}$ (right) are shown. They are
basically the same since $\zeta_d$ and $\zeta_e$ are varied in the same
range of $[-100, 100]$. And the same arguments are applied as in
the case of $g^S_{H_1\bar u u}$:
the lines with $\zeta_{d,e}=0.5$ 
are most close to $g_{_{H_1VV}}$ 
among the blue and red lines and 
those with $\zeta_{d,e}=100$ provide the envelopes which
include all the scattered points.
We see that $g^S_{H_1\bar d d}$ and $g^S_{H_1\bar e e}$ can be largely
deviated from their SM values of $1$ when $|\zeta_{d,e}|$ is large:
\begin{equation}
\left.\Delta_{H_1\bar d d\,,H_1\bar e e}
\right|_{{\rm\bf SCN200}\,,|\zeta_d|\leq 100\,,|\zeta_e|\leq 100} 
=|\zeta_{d,e}|\,\left(\frac{200~{\rm GeV}}{M_{H^\pm}}\right)^2
\lsim 1.8\, \left(\frac{1.500~{\rm TeV}}{M_{H^\pm}}\right)^2\,.
\end{equation}
Incidentally, we observe that the constraint on $|\zeta_e|$ from
the flavor-changing $\tau$ decays into light leptons 
excludes the region with 
$|g^S_{H_1\bar e e}|\gsim 60$ and $M_{H^\pm}\lsim 250$ GeV
which is not seen in the window chosen for
the scatter plot of $g^S_{H_1\bar e e}$ versus $M_{H^\pm}$
in Fig.~\ref{fig:gammch_t5},
see Eq.~(\ref{eq:upperZetae}).

\begin{figure}[t!]
\begin{center}
\includegraphics[width=8.5cm]{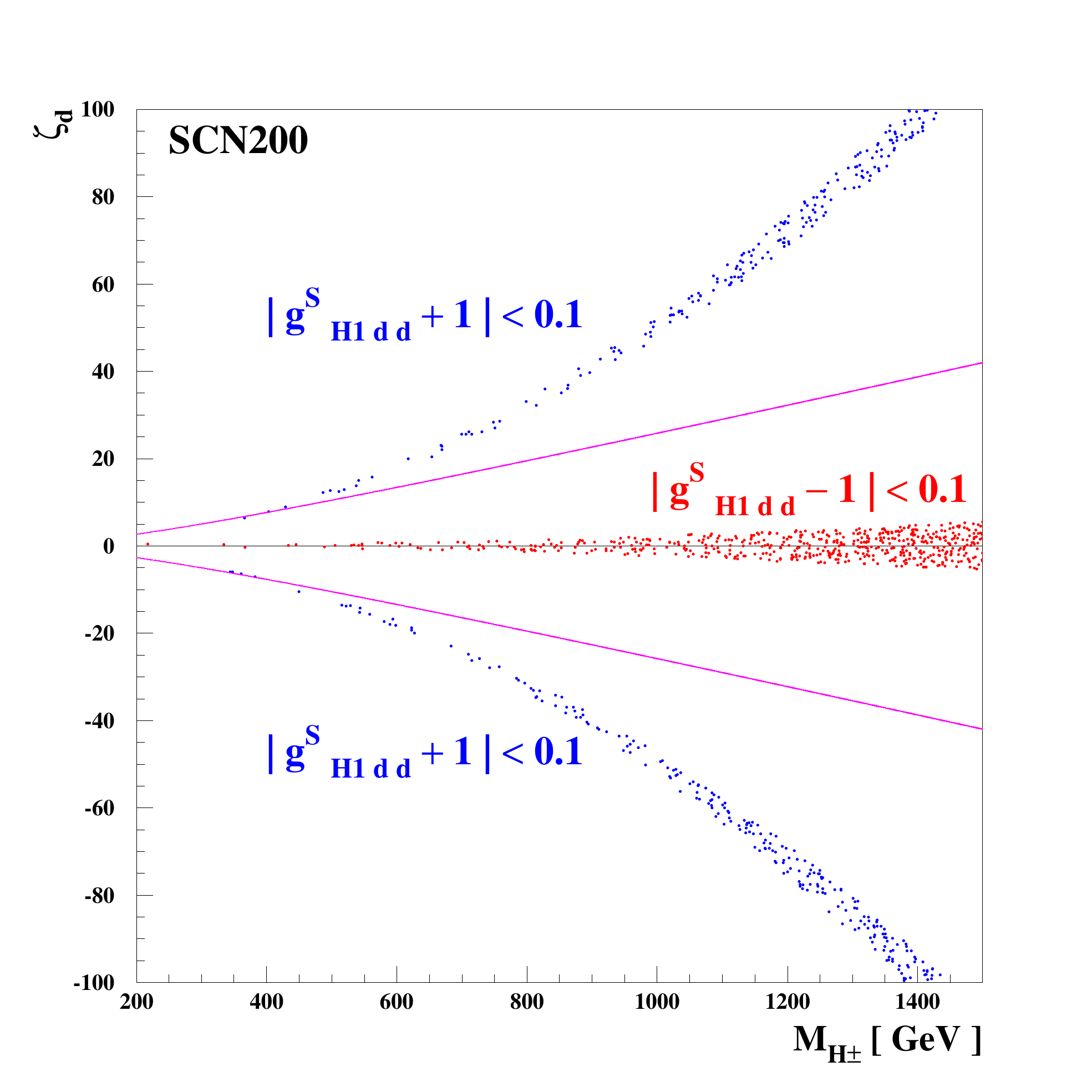}
\includegraphics[width=8.5cm]{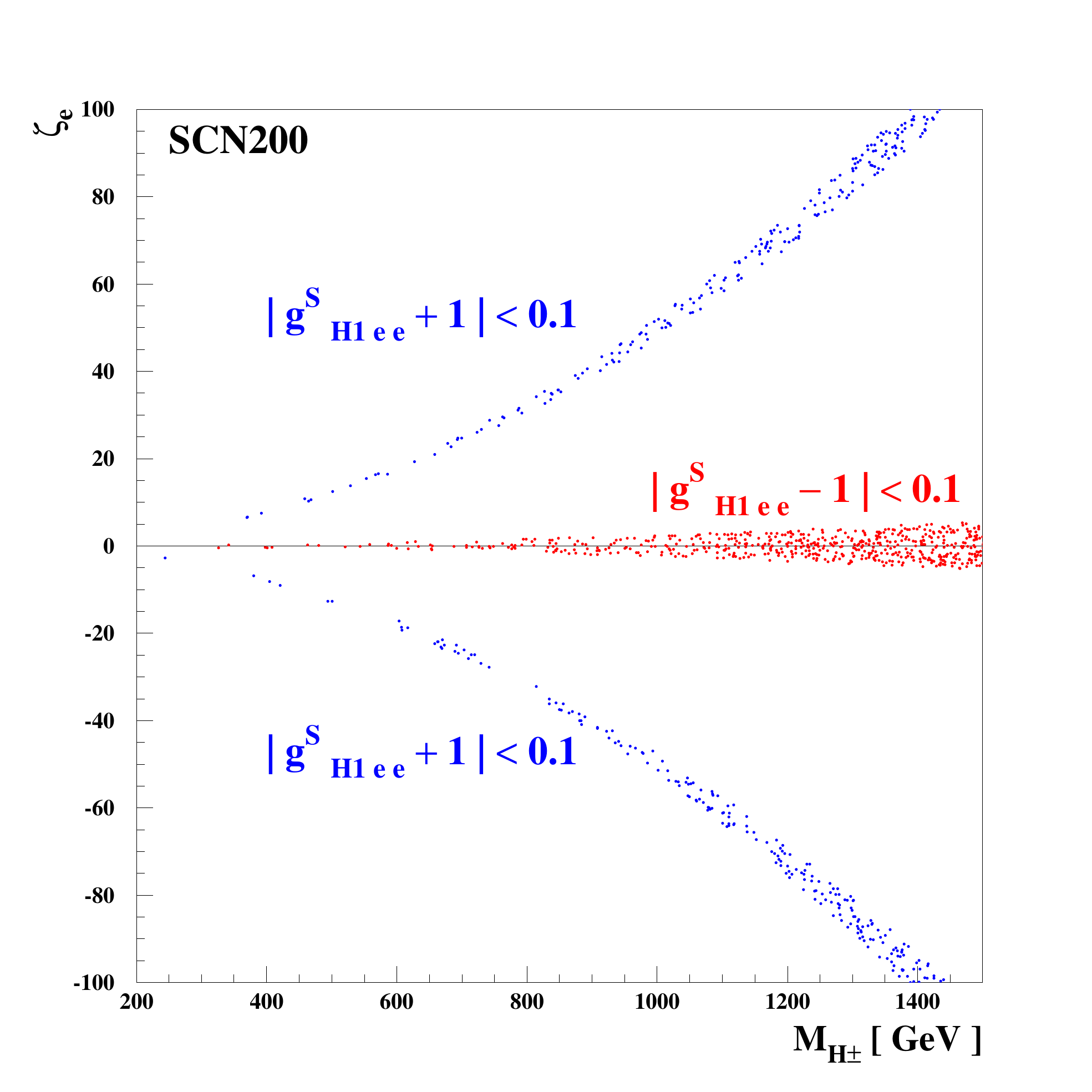}
\end{center}
\vspace{-0.5cm}
\caption{\it
Scatter plots of $\zeta_d$ versus $M_{H^\pm}$ (left) and
$\zeta_e$ versus $M_{H^\pm}$ (right) taking {\bf SCN200}.
The red and blue points denote the regions where
$g^S_{H_1\bar d d}$ and $g^S_{H_1\bar e e}$
are within the 10\% ranges of the values of $1$ and $-1$, respectively.
The magenta lines in the left panel
denote $\left|\zeta_d\right|^{\rm MIN}$
required to satisfy the
$b\to s\gamma$ constraint
through the destructive interference when
$\zeta_u\zeta_d$ is positive, see Eq.~(\ref{eq:lowerZetad}) and
the lower-left panel of Fig.~\ref{fig:x1234}.
}
\label{fig:t5zde}
\end{figure}
Of course, the alignment parameters $\zeta_{d,e}$ are constrained by
the precision LHC Higgs data. From the observation that the absolute values of
the couplings of the SM-like $H_1$ to a pair of bottom quarks and tau leptons
are required to be consistent with $1$ within
about 10\% at $1\sigma$ level~\cite{Cheung:2018ave}, one might have
$|g^S_{H_1\bar d d}\pm 1|\lsim 0.1$ and
$|g^S_{H_1\bar e e}\pm 1|\lsim 0.1$.
\footnote{The negative value of $g^S_{H_1\bar d d} \sim -1$ is less preferred than
the positive one $g^S_{H_1\bar d d} \sim +1$ at the level of
about $1.5\sigma$ considering the $b$-quark loop contributions to
the $H_1$ coupling to two gluons~\cite{Cheung:2018ave}. While, for
$g^S_{H_1\bar e e}$, the current data precision is
yet insufficient to tell its sign.  In this work, we consider
both signs for $g^S_{H_1\bar d d}$ and $g^S_{H_1\bar e e}$.}
For the positive sign, the condition
$|g^S_{H_1\bar d d}- 1| < 0.1$ constrains
$|\zeta_d|\lsim 6$, see the red points in the
left panel of Fig.~\ref{fig:t5zde}.
On the other hand $g^S_{H_1\bar d d} \sim -1$ allows the larger values of
$\zeta_d$ given by $\zeta_d=(1+\cos\gamma)/\sin\gamma
\simeq \pm 2 M_{H^\pm}^2/(200~{\rm GeV})^2$, see the blue points in the
left panel of Fig.~\ref{fig:t5zde}.
In the same panel for $\zeta_d$ versus $M_{H^\pm}$, we also show the 
lower limit on $|\zeta_d|$ from
$b\to s\gamma$ through the destructive
interference by the magenta lines, 
see Eq.~(\ref{eq:lowerZetad}) and
the lower-left panel of Fig.~\ref{fig:x1234}.
We observe that the two regions with $|g^S_{H_1\bar d d}+1|<0.1$ 
are mostly outside the band delimited by the two magenta lines implying
that large values of $|\zeta_d|$ for $g^S_{H_1\bar d d}\sim -1$
are hardly constrained by $b\to s\gamma$.
For $g^S_{H_1\bar e e}$,
the same arguments are applied,
see the right panel of Fig.~\ref{fig:t5zde}. 
Note that the constraints from
the flavor-changing $\tau$ decays into light leptons given by
Eq.~(\ref{eq:upperZetae}) are
too weak to affect those on $g^S_{H_1\bar e e}$
by the precision LHC Higgs data.

Lastly, we comment on the wrong-sign alignment limit
in the four types of conventional 2HDMs
in which the $H_1$ couplings to the down-type quarks and/or
those to the charged leptons are equal in strength but
opposite in sign to the corresponding SM ones.
The two couplings $g^S_{H_1\bar d d}$ and $g^S_{H_1\bar e e}$ are
completely independent from each other in general 2HDM.
But, in the conventional four types of 2HDMs,
they are related. We observe that the couplings
are given by either $\cos\gamma-\sin\gamma/t_\beta$ or
$\cos\gamma+t_\beta \sin\gamma$ in any type of 2HDMs,
see Table~\ref{tab:2hdmtype}.
In this case, $\cos\gamma-\sin\gamma/t_\beta=\pm 1$
for the $t_\beta$ value which
makes $\cos\gamma+t_\beta \sin\gamma=\mp 1$.
This implies that, independently of 2HDM type and
regardless of the heavy Higgs-mass scale,
all four types of 2HDMs could be viable against
the LHC Higgs precision data in the {\it wrong-sign}
alignment limit.
\section{Conclusions}
We have studied the alignment of Yukawa couplings
in the framework of general 2HDMs
identifying the lightest neutral Higgs boson as the 125 GeV one discovered
at the LHC.
We take the so-called Higgs basis
\cite{Donoghue:1978cj,Georgi:1978ri,
Botella:1994cs,Branco:1999fs,Davidson:2005cw,Haber:2006ue,Boto:2020wyf}
for the Higgs potential in which only one of the two doublets contains
the non-vanishing vacuum expectation value $v$.
For the Yukawa couplings,
rather than invoking the Glashow-Weinberg condition~\cite{Glashow:1976nt}
based on appropriately defined discrete $Z_2$ symmetries,
we require the absence of tree-level FCNCs
by assuming that the Yukawa matrices describing the couplings of the
two Higgs doublets to the SM fermions
are aligned in the flavor space
\cite{Manohar:2006ga,Pich:2009sp,Penuelas:2017ikk}.

For a numerical study, we further assume that the seven quartic couplings
$Z_{i=1-7}$ appearing in the Higgs potential and
the three alignment parameters $\zeta_{f=u,d,e}$
for Yukawa couplings are all real by anticipating that the
impacts due to CP-violating phases of $Z_{5,6,7}$ and $\zeta_{f}$'s
on the alignment of Yukawa couplings are redundant.
In this case, in addition to the vev $v$ and masses of the SM fermions,
the model can be fully described by specifying
the following set of 11 free parameters:
$$
\left\{M_{h}=M_{H_1}=125.5~{\rm GeV},M_{H},M_{A},M_{H^\pm},\gamma,Z_2,Z_3,Z_7;
\zeta_u,\zeta_d,\zeta_e\right\}\,,
$$
where $M_{h,H}$ (with $M_h<M_H$) and $M_A$ denote the masses of CP-even and CP-odd neutral
Higgs bosons, respectively, and
the mixing between the two CP-even neutral states is described by the
angle $\gamma$. The quartic couplings 
$Z_{1,4,5,6}$ are determined in terms of 
$M_{h,H,A}$, $M_{H^\pm}$, $\gamma$, and $v$. 
The quartic coupling $Z_3$ is related to the massive parameter 
$Y_2$ appearing in the Higgs potential
through $Y_2=M_{H^\pm}^2-Z_3v^2/2$.
On the other hand, the the other quartic couplings 
$Z_2$ and $Z_7$ have no direct relevance to the masses 
and mixing of Higgs bosons.
%
But, we observe that
they are interrelated with the other five quartic couplings of $Z_{1,3-6}$
through the perturbative unitarity (UNIT) conditions and
those for the Higgs potential to be bounded from below (BFB).
We note that the UNIT and BFB conditions are basis-independent, i.e.,
the same in any basis~\cite{Jurciukonis:2018skr}.
Also considered are the constraints from
the electroweak (ELW) oblique corrections to the $S$ and
$T$ parameters which are expressed in terms of the
physical observable quantities of $M_{h,H,A}$, $M_{H^\pm}$,
and $g_{_{H_iVV}}$ which are again
invariant under a change of basis~\cite{Grzadkowski:2018ohf}.
%
We further consider the constraints on
the alignment parameters $\zeta_{f=u,d,e}$ from
flavor-changing $\tau$ decays, $R_b$,
$\epsilon_K$, and
the radiative $b\to s\gamma$ decay.

\smallskip

For the independent model parameters
and the rephasing invariant combinations of CP-violating phases,
among the several points already discussed in the literature,
we highlight the following ones:
\begin{enumerate}
\item
The general 2HDM potential can be fully specified with
the masses of the charged and three neutral Higgs bosons, the 
orthogonal neutral-Higgs boson mixing matrix $O_{3\times 3}$
and the three dimensionless
quartic couplings of $Z_{2,3,7}$ in addition to the vev $v$.
\item For the CP phases, as far as the Higgs potential 
{\it and} the three complex alignment parameters for the Yukawa 
couplings are involved, 
the Lagrangians are invariant under the following
phase rotations:
\begin{eqnarray}
&&
{\cal H}_2 \to  {\rm e}^{+i\zeta}\,{\cal H}_2\,;\nonumber\\  &&
Y_3 \to Y_3\,{\rm e}^{-i\zeta}\,, \,
Z_5 \to Z_5\,{\rm e}^{-2i\zeta}\,, \,
Z_6 \to Z_6\,{\rm e}^{-i\zeta}\,, \,
Z_7 \to Z_7\,{\rm e}^{-i\zeta}\,;\nonumber \\ && \,
\zeta_u \to \zeta_u\,{\rm e}^{+i\zeta}\,,\,
\zeta_d \to \zeta_d\,{\rm e}^{-i\zeta}\,,\,
\zeta_e \to \zeta_e\,{\rm e}^{-i\zeta}\,,
\end{eqnarray}
which, taking account of the CP odd tadpole condition $Y_3+Z_6\,v^2/2=0$,
lead to the following
five rephasing-invariant CPV phases:
\begin{eqnarray}
\hspace{-1.0cm}
{\rm Arg}[Z_6 (Z_5^*)^{1/2}]\,, \,
{\rm Arg}[Z_7 (Z_5^*)^{1/2}]\,; \,
{\rm Arg}[\zeta_u (Z_5)^{1/2}]\,, \,
{\rm Arg}[\zeta_d (Z_5^*)^{1/2}]\,, \ \ {\rm and} \ \
{\rm Arg}[\zeta_e (Z_5^*)^{1/2}]\,,
\end{eqnarray}
pivoting, for example, around the complex quartic coupling $Z_5$.
\end{enumerate}
Incidentally, it is well known that
the 3 alignment parameters are
the same $\zeta_u=\zeta_d=\zeta_e=1/t_\beta$ in the type-I 2HDM.
In this case, they cannot be
significantly large than 1 since $t_\beta \ll 1$ leads to a
non-perturbative top-quark Yukawa coupling and a Landau pole
close to the TeV scale. Therefore, in the type-I model
among the 4 conventional 2HDMs, all the Yukawa couplings
of the lightest Higgs boson most
quickly approach the corresponding SM values as the masses of
the heavy neutral Higgs bosons increase
and their decouplings are least delayed.

\smallskip
We further suggest the following points 
as the main results specifically pertinent to our analysis:
\begin{enumerate}
\item
By scanning the heavy Higgs masses up to 1.5 TeV, we find that
the UNIT and BFB conditions combined with the ELW constraint
restrict the quartic couplings as follows:
\begin{eqnarray}
&& \hspace{0.33cm}
0.1\lsim Z_1 \lsim 2.0\,, \ \ \
0  \lsim Z_2 \lsim 2.1\,, \ \ \
-2.4 \lsim Z_3 \lsim 8.0\,, \ \ \
-6.3 \lsim Z_4 \lsim 6.0\,, \nonumber \\
&&
-1.9\lsim Z_5\lsim 1.6\,, \ \ \
-2.7\lsim Z_6\lsim 2.7\,, \ \ \
-2.7\lsim Z_7\lsim 2.7\,.
\end{eqnarray}
And, when $M_{H^\pm} \gsim$ 500 GeV (1 TeV), we also find that
\begin{eqnarray}
&&
|M_H-M_A|/{\rm GeV}\lsim 200\,(100) \,, \ \ \
|M_{H^\pm}-M_H|/{\rm GeV}\lsim 200\,(110) \,, \nonumber \\
&&
|M_{H^\pm}-M_A|/{\rm GeV}\lsim 200\,(110) \,, \ \ \
|\gamma|\lsim  0.8\,(0.14)\,.
\end{eqnarray}
\item
As the masses of heavy Higgs bosons increase,
compared to the $g_{_{H_1VV}}$ coupling of the lightest Higgs boson
to a pair of massive vector bosons,
the decoupling of the Yukawa couplings to the lightest Higgs boson
is delayed by the amount of 
the Yukawa delay factor
$\Delta_{H_1\bar f f}=
|\zeta_{f}|(1-g_{_{H_1VV}}^2)^{1/2}$ which is basis-independent 
and can be generally used even in the presence of CPV phases.
Therefore, though $g_{_{H_1VV}}$ 
approaches its SM value of 1 very quickly
as the masses of heavy Higgs bosons increase, the coupling of
$H_1$ to a pair of fermions can significantly
deviate from its SM value if $|\zeta_{f}|$ is large.
Note that 
$|\zeta_u|$ is constrained to be small by $R_b$ and 
$\epsilon_K$, see Eq.~(\ref{eq:upperZetau}). While $|\zeta_d|$ and $|\zeta_e|$ 
are constrained to be small
by the LHC precision Higgs data when the corresponding Yukawa couplings
are with the similar strength and the same sign as the SM ones. 
But it could be large when the Yukawa coupling
takes the wrong sign.
\item
The wrong-sign alignment, in which the $H_1$ couplings
to a pair of $f$-type fermions are equal in strength but
opposite in sign to the corresponding SM ones, occurs when
$\zeta_{f}=(1+\cos\gamma)/\sin\gamma$
independently of the heavy Higgs-boson masses.
In the conventional four types of 2HDMs,
$\zeta_f=-t_\beta$ or $1/t_\beta$ and the Yukawa couplings
are given by either $\cos\gamma-\sin\gamma/t_\beta$ or
$\cos\gamma+t_\beta \sin\gamma$ in any type of 2HDMs.
We observe that
$\cos\gamma-\sin\gamma/t_\beta=\mp 1$ for the $t_\beta$ value making
$\cos\gamma+t_\beta \sin\gamma=\pm 1$ and
any type of conventional
2HDMs is viable against the LHC Higgs precision data.
\item
Last but not least, 
by combining with the upper limit on $|\zeta_u|$ from $R_b$ and $\epsilon_K$,
we derive the {\it lower} limit on $|\zeta_d|$
independently of $\zeta_u$ and $\zeta_e$
when the non-SM contribution to $b\to s\gamma$
is about two times of the SM one at the amplitude level.

\end{enumerate}

%

%
%
\section*{Acknowledgment}
We thank Seong Youl Choi for helpful comments on the manuscript.
This work was supported by the National Research Foundation (NRF) of Korea
Grant No. NRF-2021R1A2B5B02087078 (J.S.L., J.P.).
In addition, the work of J.S.L. was supported in part by
the NRF of Korea Grant No. NRF-2022R1A5A1030700 and
the work of J.P. was supported in part by
the NRF of Korea Grant No. NRF-2018R1D1A1B07051126.
%


\end{document}